\documentclass[lettersize,journal]{IEEEtran}
\usepackage{float}
\usepackage{color}

\usepackage{amsmath,amsfonts}
\usepackage{algorithmic}
\usepackage{array}
\usepackage[caption=false,font=normalsize]{subfig}
\usepackage{textcomp}
\usepackage{stfloats}
\usepackage{url}
\usepackage{verbatim}
\usepackage{graphicx}
\usepackage{booktabs}
\usepackage{hyperref}
\usepackage{orcidlink}
\usepackage{tikz}
\def\BibTeX{{\rm B\kern-.05em{\sc i\kern-.025em b}\kern-.08em
    T\kern-.1667em\lower.7ex\hbox{E}\kern-.125emX}}
\usepackage{balance}
\usepackage{xcolor}
\usepackage[bb=boondox,bbscaled=.95,cal=boondoxo]{mathalfa} 

\begin{document}

\title{Last Week with ChatGPT: \\A Weibo Study on Social Perspective Regarding ChatGPT for Education and Beyond}

\author{Yao Tian, Chengwei Tong, Lik-Hang Lee, Reza Hadi Mogavi, Yong Liao, and Pengyuan Zhou
\thanks{Yong Liao, Chengwei Tong and Yao Tian are with the School of Cyber Science and Technology, University of Science and Technology of China, China (e-mail: \href{mailto:yliao@ustc.edu.cn}{yliao@ustc.edu.cn}, {\{\href{mailto:cwtong@mail.ustc.edu.cn}{cwtong}, \href{mailto:tyzkd@mail.ustc.edu.cn}{tyzkd}\}@mail.ustc.edu.cn).}}
\thanks{Lik-Hang Lee is with the Department of Industrial and Systems Engineering, The Hong Kong Polytechnic University, Hong Kong SAR (e-mail: \href{mailto:lik-hang.lee@polyu.edu.hk}{lik-hang.lee@polyu.edu.hk}).}
\thanks{Reza Hadi Mogavi is with the HCI Games Group (HCIGG), University of Waterloo, Canada (e-mail: \href{mailto:rhadimog@uwaterloo.ca}{rhadimog@uwaterloo.ca}).}
\thanks{Pengyuan Zhou is with the Department of Electrical and Computer Engineering, Aarhus University, Denmark (e-mail: \href{mailto:pengyuan.zhou@ece.au.dk}{pengyuan.zhou@ece.au.dk}).}
\thanks{Yong Liao is the corresponding author.}
}

\markboth{IEEE TRANSACTIONS ON LEARNING TECHNOLOGIES}%
{Yao \MakeLowercase{\textit{et al.}}: A Perspective Study on Chinese Social Media regarding LLM for Education and Beyond}

\maketitle

\begin{abstract}
The application of AI-powered tools has piqued the interest of many fields, particularly in the academic community. This study uses ChatGPT, currently the most powerful and popular AI tool, as a representative example to analyze how the Chinese public perceives the potential of large language models (LLMs) for educational and general purposes. Although facing accessibility challenges, we found that the number of discussions on ChatGPT per month is 16 times that of Ernie Bot developed by Baidu, the most popular alternative product to ChatGPT in the mainland, making ChatGPT a more suitable subject for our analysis.
The study also serves as the first effort to investigate the changes in public opinion as AI technologies become more advanced and intelligent. The analysis reveals that, upon first encounters with advanced AI that was not yet highly capable, some social media users believed that AI advancements would benefit education and society, while others feared that advanced AI, like ChatGPT, would make humans feel inferior and lead to problems such as cheating and a decline in moral principles. The majority of users remained neutral. Interestingly, with the rapid development and improvement of AI capabilities, public attitudes have tended to shift in a positive direction. We present a thorough analysis of the trending shift and a roadmap to ensure the ethical application of ChatGPT-like models in education and beyond.
\end{abstract}


\section{Introduction}

\IEEEPARstart{C}{hatbots} can understand natural language input and respond in a human-like manner, making them ideal for tasks such as answering questions, providing guidance, and even offering emotional support. In customer service, chatbots can handle common problems and provide support and guidance. For example, Cui et al.~\cite{cui2017superagent} introduced SuperAgent, which is cost-effective when answering repetitive questions, freeing up human support staff to answer more complicated questions. Chatbot can also help users find interesting products and recommend related products in online shopping services. As confirmed by \cite{han2021impact}, anthropomorphic bot plays a positive role in shaping consumers’ intentions to purchase through chatbot commerce. In the medical field, chatbots serve as medical manuals to help patients become aware of their illnesses and improve their health. For example, a text diagnosis bot enables sufferers to join in analyses of their medicinal matters and present a personalized analysis report about the symptoms~\cite{divya2018self}. Meanwhile, research on the application of chatbots to education and other areas is still in its infancy~\cite{hwang2021review} and waits for more exploration.

On November 30, 2022, OpenAI released ChatGPT~\cite{openai_chatgpt}, a large language model (LLM), which enables realistic conversations with humans, sequential inquiries, faulty dispute premises, and rejects unsuitable requests. ChatGPT can also generate original content like songs, scripts, and code, and even imitate different personas to interact with provided premises. OpenAI elevated ChatGPT's capabilities to the next level with the release of GPT-4 in March 2023~\cite{openai_gpt4}. GPT-4 is a big multimodal model that accepts image and text input and generates accurate text output. Experiments indicate that GPT-4 performs at a human level on a variety of professional and academic standards~\cite{OpenAI2023GPT4TR}.

Whether ChatGPT can pass the Turing test~\cite{turing1950computing} in the future remains uncertain, but it is expected to significantly impact various life aspects, particularly education. GPT-4 stands out with its image recognition feature, e.g., understanding and answering a physics question accompanied by an image. Microsoft has integrated GPT-4 into Microsoft 365 Copilot~\cite{microsoft_copilot}, helping users with tasks like generating content and analyzing data, thus boosting productivity. Nonetheless, there are concerns that this might lead to a decrease in human learning abilities and an increased reliance on technology.

To examine China's societal response to groundbreaking AI technologies such as ChatGPT, we formulated three primary research questions as follows:

\begin{itemize}
    \item \textbf{RQ1: What is the prevailing sentiment on Chinese social media platforms regarding revolutionary AI innovations like ChatGPT?} 

    \item \textbf{RQ2: How does the release of a more powerful AI product, GPT-4 (well-acknowledged SOTA so far), impact public attitudes toward this kind of application?}

    \item \textbf{RQ3: What concerns and issues are commonly held by the public with regard to AI-human collaboration and interaction, especially in the field of education?}
\end{itemize}

To address these research questions, we collected data by crawling posts containing ``ChatGPT'' and other relevant keywords on Sina Weibo, China's largest microblogging platform with 605 million monthly active users. We chose ChatGPT over more easily accessible AI products within China for several reasons.

Firstly, we emphasize that even though there are usage restrictions for ChatGPT in China, alternative access methods are available. Additionally, numerous online videos demonstrate the capabilities and functionalities of GPT, which help users understand and appreciate its potential.

Secondly, Our research primarily focuses on the general awareness and understanding of AI-powered tools among the public, not just those with hands-on experience. Understandings can be acquired through various sources such as media reports, academic discussions, and social networks, which are equally important and effective. This approach is common in social science research, particularly in the early stages of technological adoption such as Self-driving automobiles.

As a comparison, we also crawled discussion posts about Ernie Bot, an AI chatbot developed by Baidu, which is the most popular and representative ChatGPT alternative in China. As shown in Table \ref{tab_count}, there were only 25,462 posts about Ernie Bot from its release on 16 March 2023 to 17 May 2024, compared to 145,219 posts about ChatGPT in the five months from 30 November 2022 to 4 May 2023. The discussions about ChatGPT per month were approximately 16 times more frequent than those about Ernie Bot,  emphasizing the significant social media presence and impact of ChatGPT. Therefore, analyzing ChatGPT, a remarkable AI application, is more appropriate for our study.

\begin{table*}[!t]
    \centering
    \caption{Brief introduction of some articles and the distinctions in our work}
    \begin{tabular}{p{2.6cm}p{14.6cm}}
        \toprule
        \textbf{Papers} & \textbf{Brief Introduction} \\
        \midrule
        Mogavi et al.~\cite{mogavi2023exploring} & This article analyzed data from $\mathbb{X}$, Reddit, YouTube, and LinkedIn to explore the user experience and perspectives of early adopters toward ChatGPT in various education sectors.\\
        Baidoo-Anu et al.~\cite{baidoo2023education} & This review synthesized recent extant literature to offer some potential benefits of ChatGPT in promoting teaching and learning.\\
        Kasneci et al.~\cite{kasneci2023chatgpt} & This commentary discusses the potential benefits and challenges of LLMs in education from student and teacher perspectives.\\
        Rudolph et al.~\cite{rudolph2023chatgpt} & This review adopts a desktop analysis approach and conducts an extensive literature review, focusing on ChatGPT's implications for higher education.\\
        Zhai~\cite{zhai2022chatgpt} & This study piloted ChatGPT to write an academic paper and reflected on the potential impacts of  AI tools on education based on the user experience.\\
        Tlili et al.~\cite{tlili2023if} & This study collected and analyzed $\mathbb{X}$ data, and examined ChatGPT in education among early adopters through a qualitative instrumental case study.\\
        Kung et al.~\cite{kung2023performance} & This article evaluated the performance of ChatGPT on the United States Medical Licensing Exam and suggested that LLMs may have the potential to assist with medical education and clinical decision-making.\\
        Rahman et al.~\cite{rahman2023chatgpt} & This study analyzed data from articles, websites, blogs, and various media, finding ChatGPT effective in generating initial academic research ideas. However, it identified challenges in literature synthesis, citations, problem statements, research gaps, and data analysis, suggesting the need for guidelines on the proper use of LLMs.\\
        Bahrini et al.~\cite{bahrini2023chatgpt} & This article reviewed the existing literature, and examined the applications, opportunities, and threats of ChatGPT in 10 main domains, providing detailed examples for the business and industry as well as education.\\
        Liu et al.~\cite{liu2023summary} & This article performed an in-depth analysis of 194 relevant papers on arXiv, and the findings reveal a significant and increasing interest in ChatGPT/GPT-4 research.\\
        \textbf{Our article} & Our article differs in the following ways: 1) we collect data from Chinese social media, Weibo, which serves as a unique public opinion environment and data source of Chinese society that has not been explored; 2) we compare the people's attitudes and opinions before and after the release of GPT-4; 3) we expand the investigation to more interactions scenarios beyond education.\\
        \bottomrule
    \end{tabular}
    \label{tab_compare}
    \vspace{-4mm}
\end{table*}

After getting the data, we categorized the posts gathered by the specified keywords into two major groups: AI for education and human-AI interaction. Each major category includes three scenarios. We conducted sentiment and topic analyses on these posts to assess the positive and negative impacts of ChatGPT, aiming to enhance its benefits while mitigating potential drawbacks for its safe and responsible use.

\textbf{After extensive analysis, we found the following key takeaways:
\begin{itemize}
    \item On Chinese social media, most people hold a neutral stance towards the use of ChatGPT in education, with more negative than positive attitudes. 
    \item The launch of GPT-4 has improved public perception of AI applications like ChatGPT, although there remains a cautious approach towards such technologies.
    \item Public concerns include privacy infringement, content bias, inaccuracy, harm to educational fairness, and over-reliance on AI applications. 
\end{itemize}
}

The remainder of this paper is organized as follows. Related works are discussed in Section~\ref{sec:related}. Section~\ref{sec:method} presents our data collection strategy and statistics, and analysis tools. Our experimental results and analysis are illustrated in Section~\ref{sec:results}. Section~\ref{sec:discussion} compares the differences between our results and those of other studies and analyses possible reasons. Finally, we provide a conclusion in Section~\ref{sec:conclusion}.

\section{Related Work}\label{sec:related}

Previous studies have explored various aspects of the application of ChatGPT in different fields, and their findings provide valuable insights into the potential benefits and challenges of this technology. 

Education is one of the most widely engaging application fields for ChatGPT, with a broad audience and easy deployment. Mogavi et al.~\cite{mogavi2023exploring} analyzed social media data from $\mathbb{X}$, Reddit, YouTube, and LinkedIn, revealing that the public discourse is generally positive, and there is enthusiasm regarding ChatGPT's use in education. Nevertheless, they found that an overreliance on AI systems may encourage superficial learning habits and diminish students' social and critical thinking skills. Baidoo-Anu et al.~\cite{baidoo2023education} highlight ChatGPT's capability to assess students' learning styles, offer personalized tutoring and feedback, and even grade assignments, thereby freeing up teachers for other tasks. However, concerns about the absence of human interaction in such generative models are noted, especially for students needing personal teacher engagement for an optimal learning experience. Kasneci et al.~\cite{kasneci2023chatgpt} believe that ChatGPT can provide customized services for learners at different stages and fields, but customizing models to specific needs, addressing bias in specific use cases, and dealing with ethical issues and copyright issues require multi-disciplinary evidence-based research and evaluation. Rudolph et al.~\cite{rudolph2023chatgpt} suggest exercising caution in adopting a regulatory approach primarily focused on uncovering academic misconduct, such as detecting the use of ChatGPT. Instead, they advocate for cultivating trusting relationships with students in a student-centric pedagogy. Zhai~\cite{zhai2022chatgpt} showed that ChatGPT could help researchers write coherent, informative, and systematic papers, and suggests adjusting learning goals—students should be able to use AI tools for scientific research, improving creativity and critical thinking, rather than just focusing on general skills.

\begin{figure*}[!ht]
\vspace{-3mm}
\centering
\includegraphics[width=6.8in]{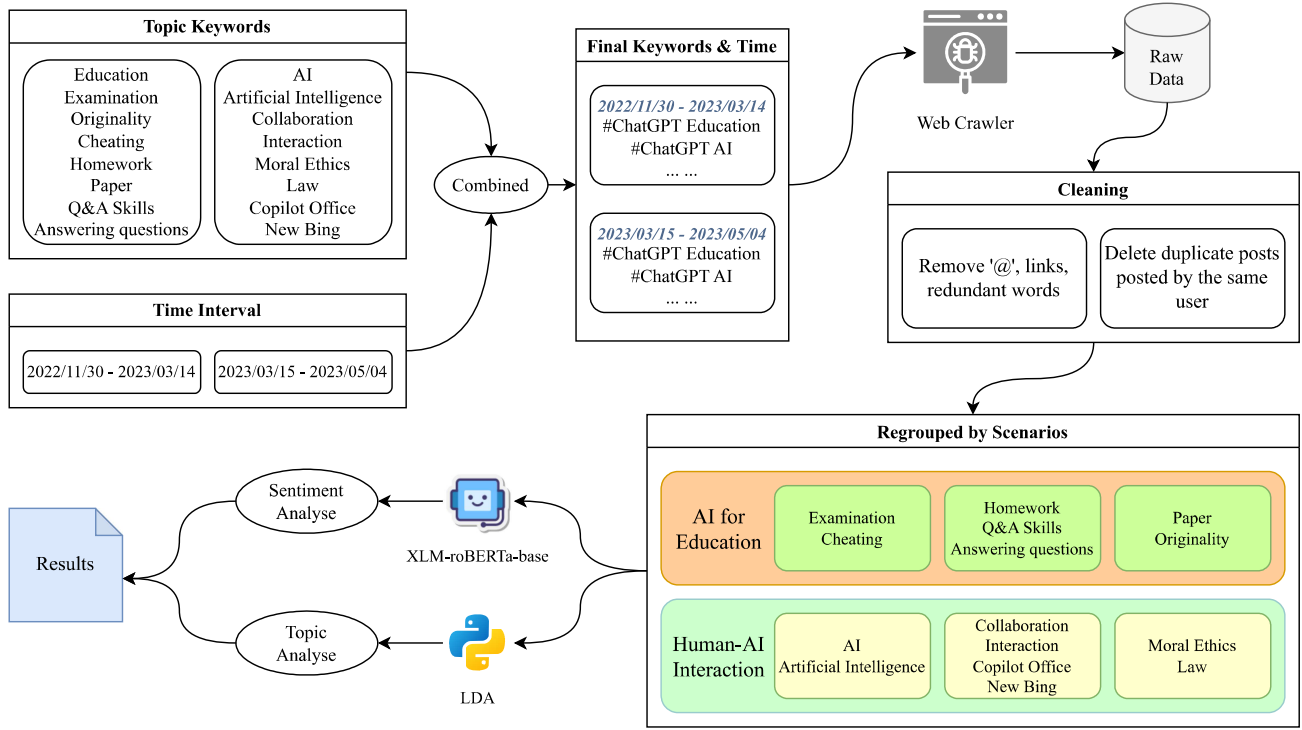}
\caption{\label{fig_flowchart}The steps for obtaining and analyzing data.}
\vspace{-5mm}
\end{figure*}

Apart from education, the application values and prospects of ChatGPT and other chatbots in different fields have also been investigated. Bahrini et al.~\cite{bahrini2023chatgpt} have examined ChatGPT's applications, opportunities, and threats in 10 main domains, providing detailed examples in business, industry and education, while Liu et al.~\cite{liu2023summary} provided insights into ChatGPT’s capabilities, potential impacts, ethical concerns, and offer direction for future advancements. Extending the focus beyond ChatGPT, the study of human-computer collaboration and relationships with AI has been a subject of interest. Skjuve et al.~\cite{skjuve2022longitudinal} conducted 12-week interviews with 25 participants to study the development of human-chatbot relationships (HCRs) and discovered HCRs evolved gradually, aligning mostly with Social Penetration Theory~\cite{carpenter2015social}. Lee et al.~\cite{lee2023exploring} found that chatbot-based social contact has promising potential for mitigating mental illness stigma and provides suggestions on how human-AI interactions can be designed to promote positive social impacts.

Meanwhile, we find that there is a gap in the comprehensive study of the opinions of Chinese society, the 2nd largest AI market\footnote{\url{https://www.idc.com/solutions/data-analytics/spending-guide}}, regarding ChatGPT and similar models. Furthermore, the impact of GPT-4 on the public's perception of ChatGPT and its influence on societal viewpoints towards such models remains under-explored. Drawing upon our data and relevant references~\cite{chou2018pursuit,avellan2020ai,kinnula2021researchers}, we present several recommendations to enhance AI inclusiveness, covering aspects such as data sources, developer teams, and user communities. These recommendations are intended to ensure equitable participation and benefits for all individuals in the AI domain.

Generally, the distinctions between our article and other comparable articles are presented in Table \ref{tab_compare}.

\section{Methods}\label{sec:method}

This study explores the potential of ChatGPT from various perspectives by analyzing individuals' attitudes and perceptions of its usage on social media platforms. Figure \ref{fig_flowchart} provides a concise overview of the steps involved in obtaining and analyzing our data.

To initiate our analysis, we selected key topic keywords in two main areas: \textit{AI for Education}, and \textit{Human-AI Interaction}, as illustrated in Figure \ref{fig_flowchart} under ``Topic Keywords.'' To achieve a comprehensive analysis, we implemented temporal segmentation, dividing the timeline into two distinct periods. The first period covers discussions from the release date of ChatGPT-3 (November 30, 2022) to the eve of GPT-4's release (March 14, 2023) and the second period begins from the release of GPT-4 (March 15, 2023) to May 4, 2023. This approach not only enabled us to systematically categorize the collected data but also to trace the evolution of public perception in response to the advancements in AI technology. The WeiboSuperSpider~\cite{WeiboSuperSpider} was instrumental in this process, allowing us to efficiently filter and collect posts that matched our designated topic keywords over the specified time frames. By combining these keywords with the \#ChatGPT hashtag as our search parameters, we were able to specifically target and analyze the discourse surrounding the integration of AI in education and human-AI interaction on Weibo.

Our decision to select Weibo as the source for our study stems from the various limitations of alternative platforms. For example, the content on WeChat Moments is more private and only accessible to added friends, which limits the scope of data collection. Xiaohongshu, despite its high daily active user count, has an uneven distribution of users in terms of gender and age, with a majority being female and content predominantly image-based. Similarly, Tiktok primarily features video content, both of which complicate text analysis. Zhihu, as a professional Q\&A platform, caters to a more specialized audience, potentially not capturing a broad public perspective. Consequently, Weibo was our preferred choice.

\begin{table}[!t]
    \centering
    \caption{The number of posts collected after cleaning}
    \begin{tabular}{p{3.7cm}p{1.05cm}p{1.05cm}p{1.16cm}}
        \toprule
        \textbf{Keywords} & \textbf{ChatGPT} & \textbf{ChatGPT} & \textbf{Ernie Bot} \\
        & 22/11/30-23/03/14 & 23/03/15-23/05/04 & 23/03/16-24/05/17\\
        \midrule
        AI & 29501 & 23547 & 15617  \\
        Artificial Intelligence & 35050 & 16471 & 6287  \\
        Paper & 8151 & 3190 & 730  \\
        Education & 6076 & 2951 & 1091  \\
        New Bing & - & 2111 & -  \\
        Law & 1658 & 1929 & 569  \\
        Homework & 4987 & 1485 & 390  \\
        Examination & 1921 & 1325 & 470  \\
        Copilot Office & - & 998 & -  \\
        Human-Computer Interaction & 957 & 299 & 140  \\
        Cheating & 1569 & 138 & 39  \\
        Moral Ethics & 204 & 129 & 4  \\
        Answering questions & 119 & 80 & 96  \\
        Q$\&$A Skills & 63 & 25 & 0 \\
        Originality & 147 & 47 & 13  \\
        Human-Computer Collaboration & 73 & 18 & 16  \\
        \bottomrule
    \end{tabular}
    \label{tab_datasize}
    \vspace{-2mm}
\end{table}

After collecting the raw data, we first cleaned it by removing common post elements like user mentions (@), links, and redundant words such as ``Collapse'', often found in lengthy Weibo posts. The number of posts after cleaning is displayed in Table \ref{tab_datasize}. We then regrouped the corresponding keywords and categorized them into our carefully selected scenarios (see Figure \ref{fig_flowchart} and Section~\ref{sec:results}) base on the following reasons:
\begin{itemize}
    \item Through our investigation of ChatGPT-enhanced learning tools\cite{gpt_academic,chatpaper,chatresponse} and analysis of collected posts, we found that the primary educational uses of ChatGPT, which garner significant attention, are related to examinations, essay writing, and homework assistance. These uses are associated with concerns regarding fairness, academic honesty, and independence in learning.
    \item Human-computer collaboration and interaction represent crucial real-world applications of AI. Research on the synergy between AI systems and humans can improve productivity, decision efficiency, and overall work performance, aligning with future trends in AI. Meanwhile, ethical and moral issues are bound to become increasingly pressing as AI extends into some morally ambiguous areas. This exploration can provide valuable reference points for addressing numerous societal and ethical challenges that AI systems may potentially introduce, such as privacy, inequality and responsibility.
\end{itemize}

This approach allowed us to explore the benefits, drawbacks, and potential risks of ChatGPT in different scenarios with finer granularity, offering richer insights than a single-keyword analysis might provide.

Subsequently, we conducted sentiment analysis on each Weibo post, using the multilingual XLM-roBERTa-base model~\cite{barbieri2022xlm} capable of processing Chinese text. The primary purpose of choosing sentiment analysis is that, as a quantitative tool, it enables us to systematically categorize and analyze subjective information within extensive text data. This approach allows us to methodically assess the emotions expressed in Weibo posts.
We opted for a three-category sentiment analysis \{positive, negative, neutral\} instead of a binary classification \{positive, negative\}, because we found many social media posts do not convey a clear emotional direction, and a binary classification could fail to accurately reflect the true sentiment of the posts. 
We then performed a chi-square test \cite{Pearson1900} on the sentiment analysis results to identify any significant changes in sentiment before and after the launch of GPT-4, shedding light on its impact on public opinion.
Additionally, we utilized the LDA model~\cite{blei2003latent} to analyze which topics people were concerned about. To effectively analyze attitudes and perspectives, we retained adjectives and adjectival nouns—words that can convey evaluative meaning. Finally, we combined these results with corresponding Weibo content to provide a comprehensive overview of general attitudes and perspectives on the various scenarios.

In our study, all data acquisition and analysis were conducted solely in Chinese to focus the study on the native opinions, encompassing everything from keywords to posts, with translation reserved for presenting results in this paper. In the process of translating vocabulary for our LDA analysis, we meticulously translated each term independently and then performed a comprehensive cross-verification. Wherever there was uncertainty, we referred to GPT to confirm the translation's authenticity, thereby ensuring the precision of our vocabulary translation.

\begin{figure*}[!t]
\vspace{-3mm}
\centering
\includegraphics[width=6.5in]{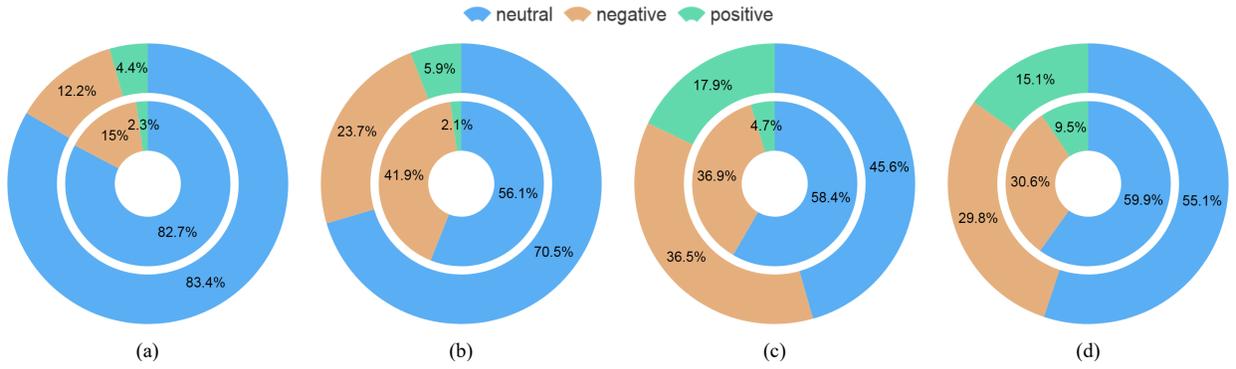}
\caption{Sentiment analysis, the inner and outer circles indicate results before and after the release of GPT-4, respectively. (a) ChatGPT and Education. (b) Scenario 1: Examination and Cheating. (c) Scenario 2: Assisting with homework. (d) Scenario 3-Paper and Originality.}
\label{fig_senti_edu}
\vspace{-3mm}
\end{figure*}

\begin{table*}[!t]
    \centering
    \caption{Chi-square Test Results of Sentiment Expression About ChatGPT and Education}
    \begin{tabular}{p{1cm} *{4}{p{0.75cm}p{1.9cm}}}
        \toprule
        \textbf{Sentiment} & \multicolumn{2}{p{2.7cm}}{\textbf{Education}} & \multicolumn{2}{p{2.7cm}}{\textbf{Examination and Cheating}} & \multicolumn{2}{p{2.7cm}}{\textbf{Assisting with Homework}} & \multicolumn{2}{p{2.7cm}}{\textbf{Paper and Originality}} \\
        \cmidrule(lr){2-3} \cmidrule(lr){4-5} \cmidrule(lr){6-7} \cmidrule(lr){8-9} 
         & $\chi^2$ & p & $\chi^2$ & p & $\chi^2$ & p & $\chi^2$ & p \\
        \midrule
        Positive & $32.15$ & $\mathbf{1.43 \times 10^{-8}}$ & $47.36$ & $\mathbf{5.90 \times 10^{-12}}$ & $288.41$ & $\mathbf{1.11 \times 10^{-64}}$ & $73.51$ & $\mathbf{1.00 \times 10^{-17}}$ \\
        Neutral & $0.51$ & $4.76 \times 10^{-1}$ & $88.68$ & $\mathbf{4.65 \times 10^{-21}}$ & $80.22$ & $\mathbf{3.36 \times 10^{-19}}$ & $21.91$ & $\mathbf{2.86 \times 10^{-6}}$ \\
        Negative & $12.73$ & $\mathbf{3.60 \times 10^{-4}}$ & $146.77$ & $\mathbf{8.81 \times 10^{-34}}$ & $0.04$ & $8.44 \times 10^{-1}$ & $0.66$ & $4.15 \times 10^{-1}$ \\
        \bottomrule
    \end{tabular}
    \label{tab_chi_square_edu}
    \vspace{-4mm}
\end{table*}

By examining users' attitudes and comments towards ChatGPT, this study can help professionals and policymakers who use or plan to use this technology understand the strengths and limitations of ChatGPT and make informed decisions.

\section{Results and Analysis}\label{sec:results}
This section provides an in-depth analysis and discussion of the collected data.

We initially counted the posts and deduplicated user statistics from our scraped data, as detailed in Table \ref{tab_count}. The data volume meets our research needs and indicates that the usage restrictions on ChatGPT in China have not significantly hindered its discussions on social media platforms (other platforms also have related discussions). Additionally, it's worth mentioning that 5,780 users discussed both ChatGPT-3 and GPT-4, suggesting that approximately 20,000 users might have been attracted by GPT-4 directly without previous releases. This highlights the fact that the release of GPT-4 attracted the attention of a significant number of Chinese users who were either unfamiliar with it or had limited prior exposure, indicating the appeal of this innovative AI tool.

Next, we analyze the results systematically by delving into six specific scenarios to dig deeper into opinions and attitudes about ChatGPT and explore why people hold different views.

\begin{table}[!h]
    \vspace{-3mm}
    \centering
    \caption{The number of posts and unique users.}
    \begin{tabular}{p{2.3cm}p{1.4cm}p{1.4cm}}
        \toprule
         & \textbf{Users} & \textbf{Posts} \\
        \midrule
        ChatGPT Period 1 & 30491 & 90476 \\
        ChatGPT Period 2 & 24647 & 54743 \\
        ChatGPT Total & 49358 & 145219 \\
        Ernie Bot & 10619 & 25462 \\
        \bottomrule
    \end{tabular}
    \label{tab_count}
    \vspace{-3mm}
\end{table}

\begin{figure*}[!t]
\centering
\subfloat[]{\includegraphics[width=3.25in]{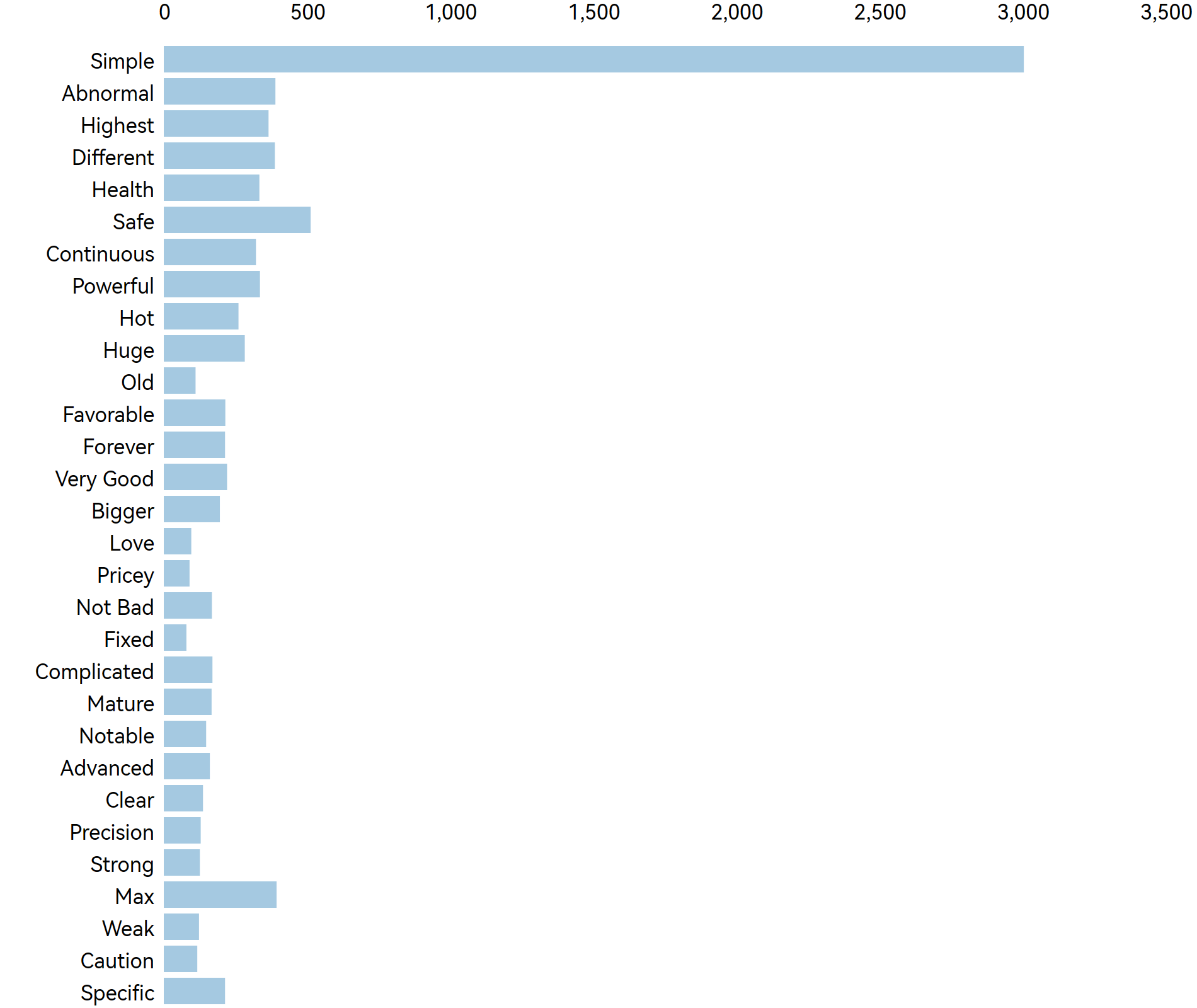}%
\label{fig_topic_edu_before}}
\hfil
\subfloat[]{\includegraphics[width=3.25in]{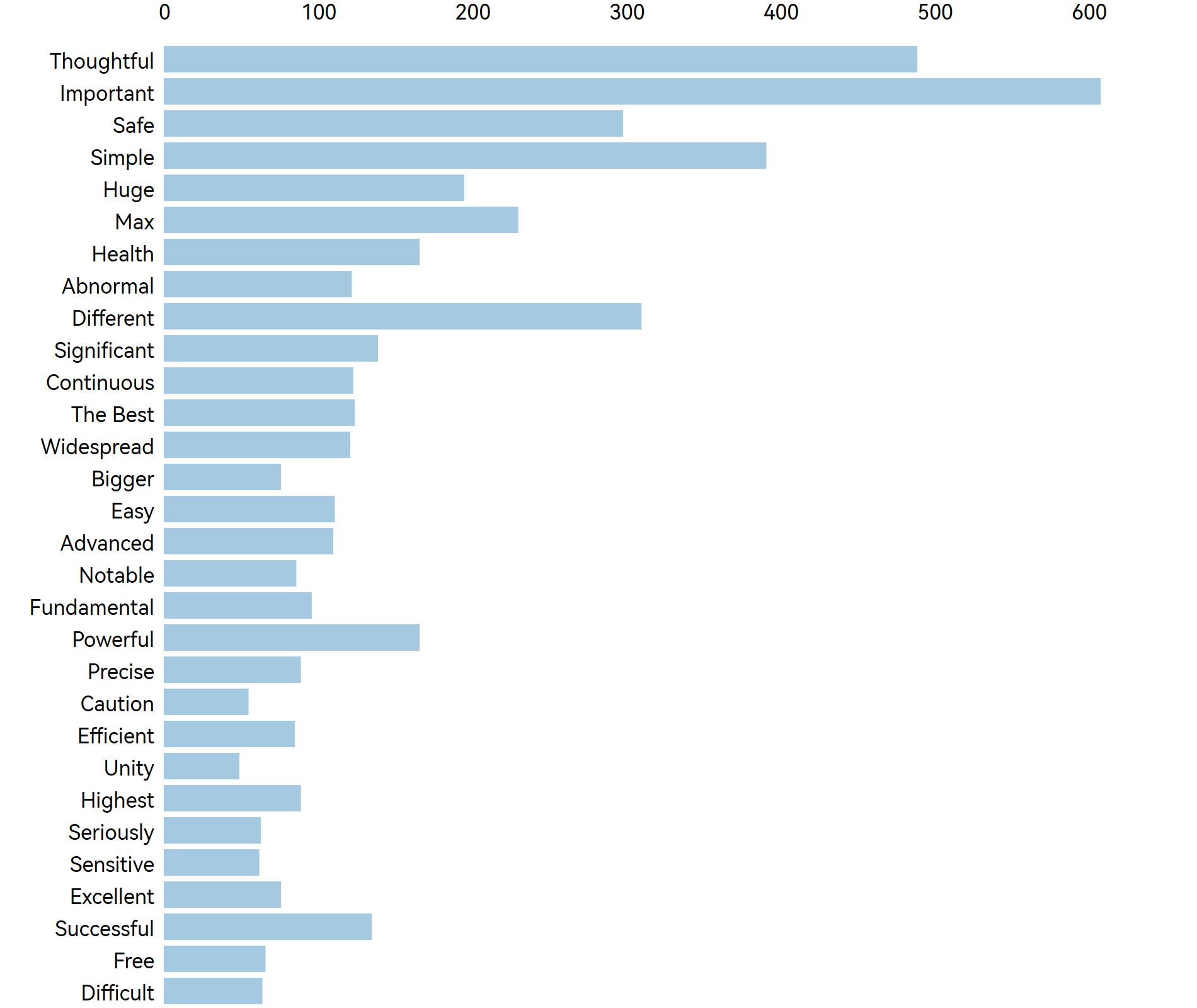}%
\label{fig_topic_edu_after}}
\caption{Topic analysis, top-20 most salient terms of ``ChatGPT and Education''. (a) before the release. (b) after the release.}
\label{fig_topic_edu}
\vspace{-3mm}
\end{figure*}

\subsection{ChatGPT and Education}

We conducted an overall sentiment analysis of Weibo posts regarding ``ChatGPT'' and ``education''. The results show that most people are hesitant about using ChatGPT in education, with 82.7$\%$ of individuals holding a neutral attitude. This highlights that the potential of ChatGPT's application in education is limited by social awareness and understanding. Meanwhile, we observe that some individuals negatively perceive ChatGPT's implementation in education, accounting for 15$\%$ of the sample. Their concerns include the fear of ChatGPT replacing the role of human teachers or stifling students' creativity. Only 2.3$\%$ of the individuals hold a positive attitude towards ChatGPT's implementation in education, likely due to the early stages of research and application in this field.

We notice a near doubling of the proportion of positive attitudes following GPT-4's release, reaching 4.4$\%$, while the proportion of negative attitudes decreases slightly to 12.2$\%$. 
In the chi-square test of the education sector and its three sub-sectors (Table~\ref{tab_chi_square_edu}), most results showed high chi-square statistics and p-values significantly below 0.05. This indicates that there is a statistically significant difference in the distribution of public sentiment on Weibo before and after the release of GPT-4.
This may be because the improvement of GPT-4 over its predecessor have enhanced people's understanding of ChatGPT's potential benefits and limitations, leading to increased confidence and optimism among users. Those initially skeptical or hesitant about using ChatGPT might have become more receptive to its application in education, reflecting this shift in attitude.

Topic analysis of Weibo posts discussing ChatGPT and education revealed a much more frequent appearance of the term ``simple'', suggesting a public expectation for ChatGPT to streamline and simplify education. Traditional education models often demand considerable time and effort from students for knowledge acquisition, while ChatGPT offers the potential for more efficient and tailored learning guidance.
Following the release of GPT-4, the distribution of topic words became more balanced, with ``important'', ``thoughtful'', and ``different'' becoming high-frequency words. 
This may indicate that people are increasingly recognizing the need to carefully consider different aspects and potential issues when integrating ChatGPT into educational practices, and are calling for well-thought-out implementation strategies. 
Keywords such as 'safety' and 'health' appeared frequently both before and after the release of GPT-4, suggesting that no matter how advanced the technology is, safety and health issues are always a focal point of concern.

\textit{Overall, the potential of using ChatGPT in education is not yet fully recognized. Next, we explored three typical education scenarios.}

\begin{figure*}[!t]
\centering
\subfloat[]{\includegraphics[width=3.25in]{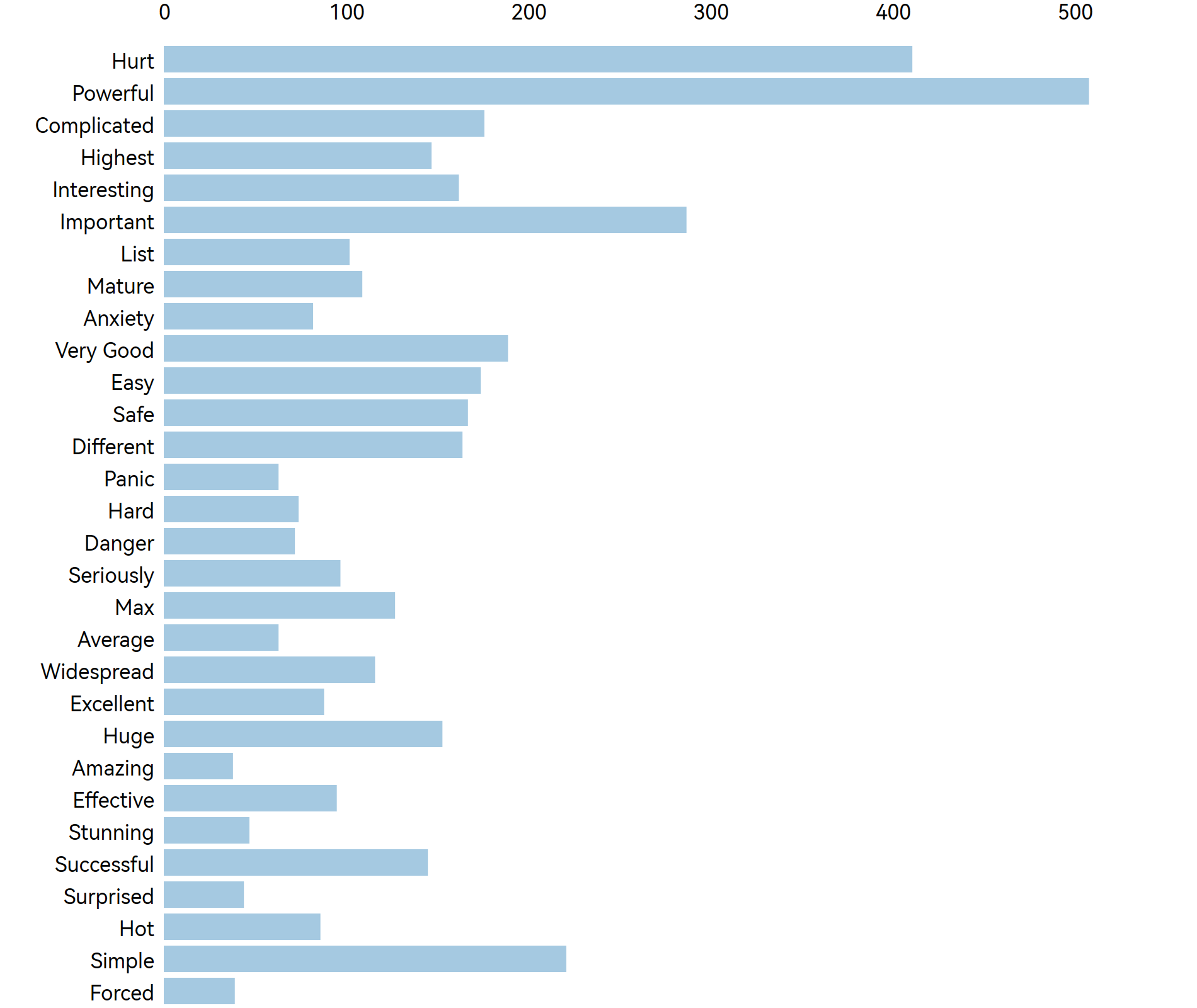}%
\label{fig_topic_exam_before}}
\hfil
\subfloat[]{\includegraphics[width=3.25in]{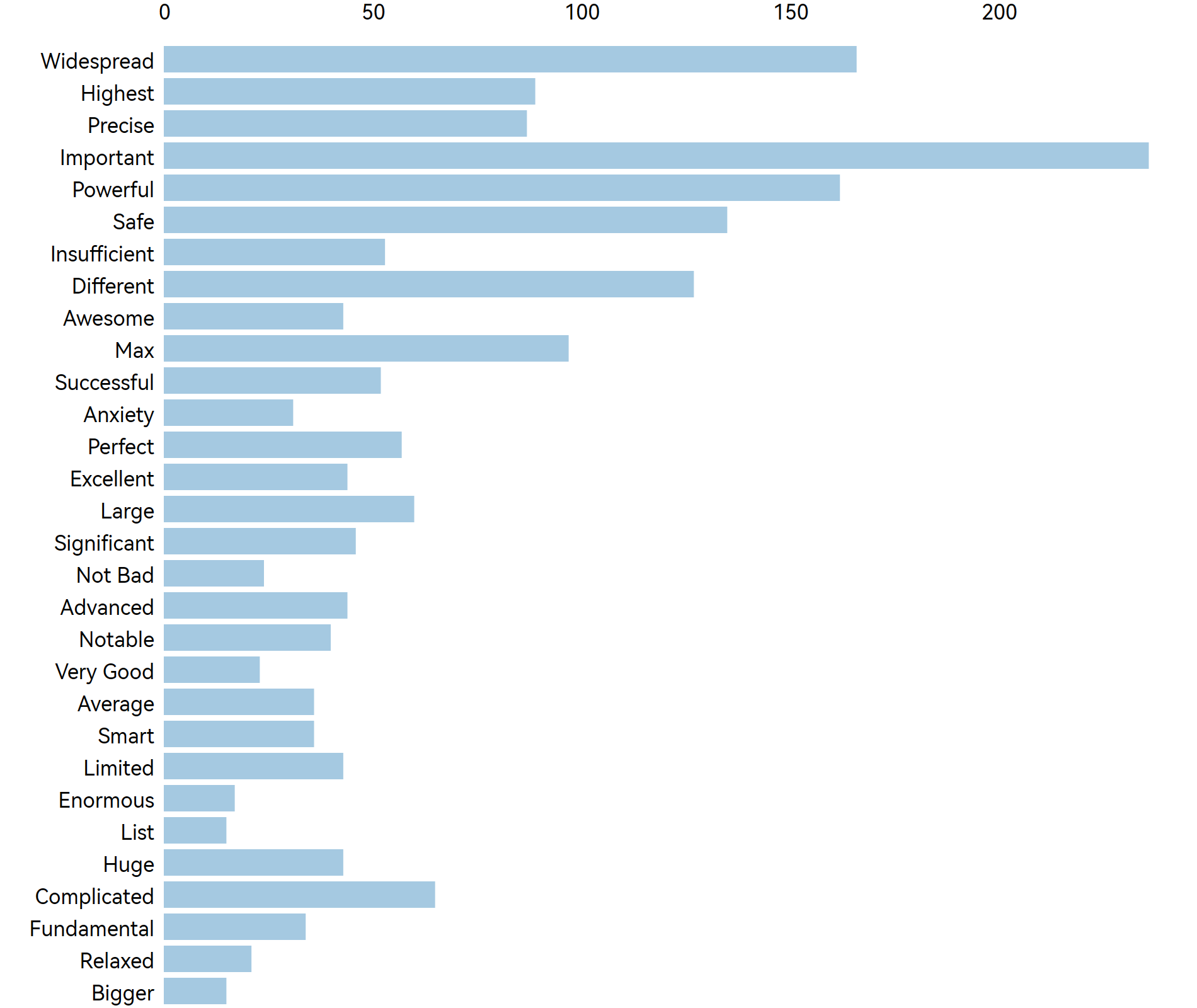}%
\label{fig_topic_exam_after}}
\caption{Topic analysis, top-20 most salient terms of ``Examinations and Cheating''. (a) before the release. (b) after the release.}
\label{fig_topic_exam}
\vspace{-4.5mm}
\end{figure*}

\subsubsection{\textbf{Scenario 1. Examination and Cheating}}

ChatGPT can generate smooth and clear answers, making it a helpful tool for exams. GPT-4 has even scored in the top 10$\%$ on a simulated lawyer exam, showcasing its human-like performance~\cite{OpenAI2023GPT4TR}. However, this also raises concerns about potential cheating risks.
In exam and cheating scenario, the proportion of individuals who view ChatGPT negatively is significantly higher than those with a positive attitude, reaching 41.9$\%$ before the release of GPT-4. This indicates that people are extremely worried that ChatGPT may be used for cheating and undermine the fairness of exams. Following the release of GPT-4, this proportion dropped to 23.7$\%$, and the proportion of positive attitudes increased from 2.1$\%$ to 5.9$\%$, Following the release of GPT-4, this proportion dropped to 23.7\%, and the proportion of positive attitudes increased from 2.1\% to 5.9\%, which is confirmed by the chi-square test results, with \(p = 1.43e-08\) for positive sentiment and \(p = 3.60e-04\) for negative sentiment. This might be attributed to the increased familiarity with ChatGPT, coupled with schools \textit{beginning to implement countermeasures like detection tools to curtail cheating}, thereby alleviating concerns and positively influencing attitudes.

The results of the topic analysis further support the differential attitudes of netizens towards the use of ChatGPT in exams. While it is commonly believed that ChatGPT can improve students' test scores by facilitating exam preparation and reviewing sessions, as evidenced by high-frequency positive terms such as ``powerful'' and ``important'' both before and after the release of GPT-4, there are also apprehensions about the potential negative implications of ChatGPT utilization. Indeed, the concerns raised by high-frequency negative terms, including ``hurt'', ``complicated'', and ``anxiety'', indicate a sense of uncertainty and anxiety surrounding the use of ChatGPT for academic purposes, with fears about potential risks to examination fairness and students' integrity.

We have exemplified some points of view in Table \ref{tab_exam} to show that ChatGPT is a double-edged sword. Thus, the use of ChatGPT in exams set possesses both merits and demerits, necessitating more comprehensive evaluations of its potential impact and role.

\begin{table}[!ht]
\centering
\caption{\label{tab_exam}Comments on ChatGPT in exam and cheating scenarios.}
\begin{tabular}{p{1.0cm}p{6.9cm}}
    \toprule
    \textbf{Sentiment} & \textbf{Content} \\
    \midrule
    \textit{Negative} & In the new semester, teachers have started to get a headache about how to prevent students from cheating with ChatGPT writing assignments and reports, which is too difficult.\\
    \textit{Neutral} & I just want to know if I can use ChatGPT to answer during the exam.\\
    \textit{Positive} & \#ChatGPT\# used 1 month to help my child's English score improve by a large margin and the test was very easy.\\
    \bottomrule
\end{tabular}
\vspace{-2mm}
\end{table}

\begin{figure*}[!t]
\centering
\subfloat[]{\includegraphics[width=3.25in]{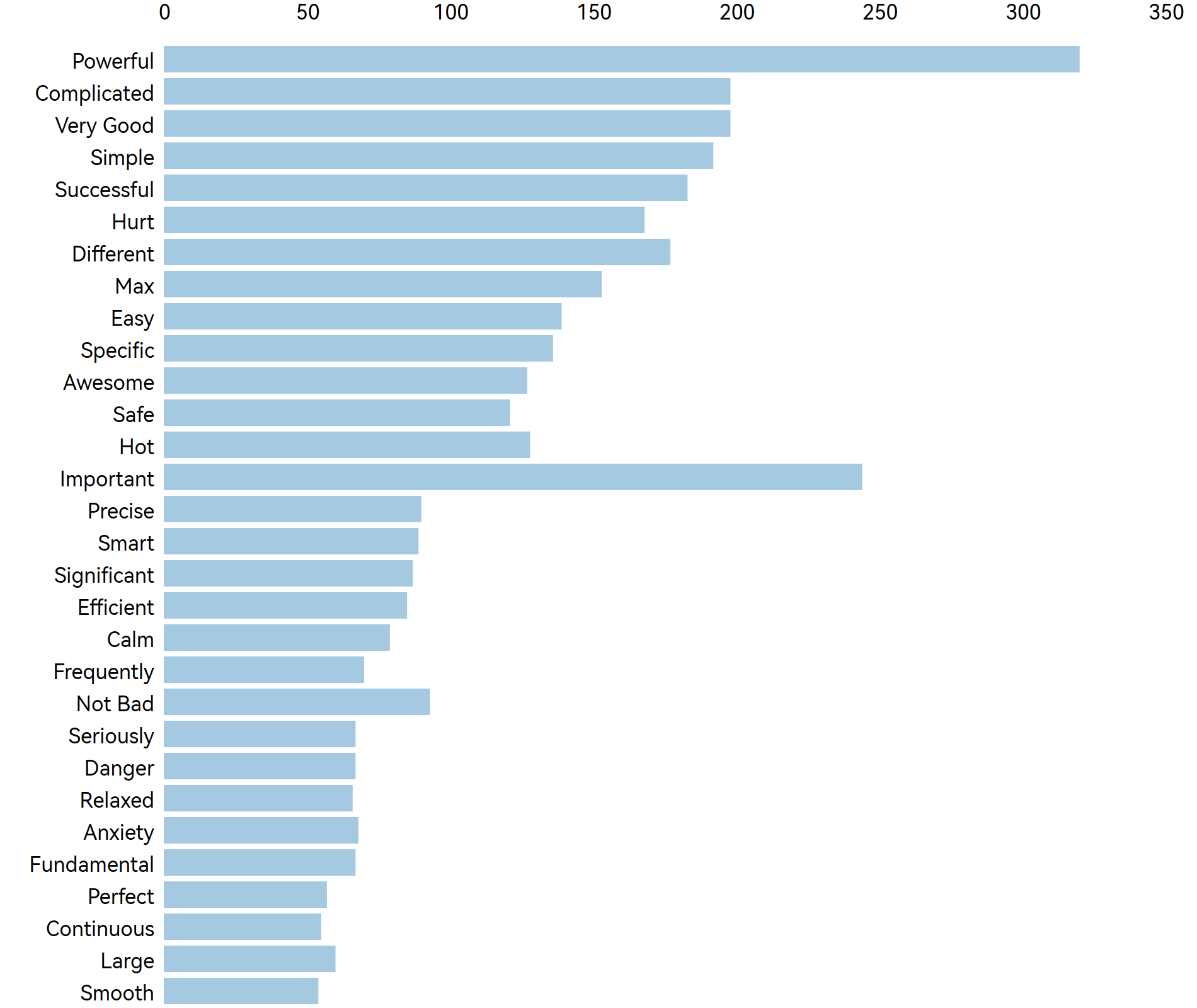}%
\label{fig_topic_homework_before}}
\hfil
\subfloat[]{\includegraphics[width=3.25in]{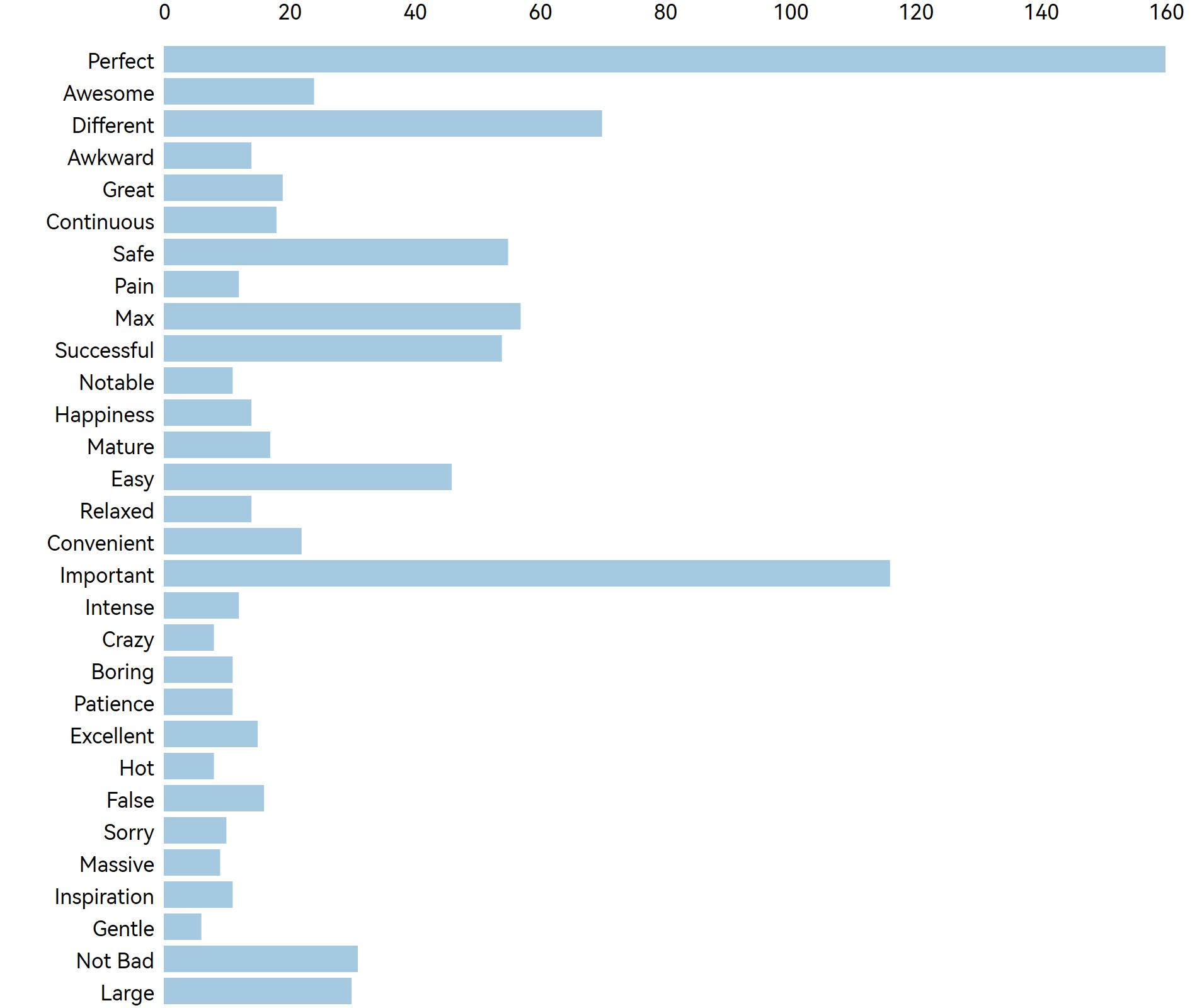}%
\label{fig_topic_homework_after}}
\caption{Topic analysis, top-20 Most Salients Terms of ``Assisting with homework''. (a) Before the Release. (b) After the Release.}
\label{fig_topic_homework}
\vspace{-4.5mm}
\end{figure*}

\subsubsection{\textbf{Scenario 2. Assisting with Homework}}

ChatGPT can play the role of a virtual teacher, helping students understand assignment requirements, improving writing skills, and even providing direct answers. Regarding using ChatGPT for homework assistance, it is worth noting that there has been little change in the proportion of negative attitudes following the release of GPT-4. However, the proportion of positive attitudes increased from 4.7$\%$ to 17.9$\%$, a growth rate that exceeds all other scenarios. 
The high chi-squared statistics and low p-values across the three sentiments also corroborate this significant change.
We believe that this might be due to the fact that using ChatGPT to assist with homework is the most direct and common scenario in which ChatGPT is applied in education. The functionality enhancement of GPT-4, combined with the experience accumulated by users in previous use, has enabled students to have a better experience when using ChatGPT to assist with their homework, thus increasing their confidence and positive evaluations of it. In addition, positive word-of-mouth and recommendations from peers may have a positive impact on students' attitudes.

Similar to the previous exam scenario, in this scenario, the keywords that contribute the most to the topic, such as ``powerful'', ``perfect'', and ``successful'', represent expectations and recognition of using ChatGPT to solve homework and improve learning efficiency. Terms like ``complicated'', ``danger'', and ``hurt'' indicate concerns about the potential harms of using ChatGPT for homework purposes. It is worth noting that such negative words have significantly decreased on Weibo after GPT-4's release, corresponding to the significant increase in the proportion of positive attitudes mentioned above.

\begin{table}[!ht]
\centering
\caption{\label{tab_homework}Comments on ChatGPT in assisting with homework scenarios.}
\begin{tabular}{p{1.0cm}p{6.9cm}}
\toprule
\textbf{Sentiment} & \textbf{Content} \\\midrule
\textit{Negative} & This software really achieves the ultimate in AI. It is too intelligent, and long-term use will naturally reduce people's thinking and creativity. If the information provided by ChatGPT is wrong, it will mislead the direction of public opinion. \\
\textit{Neutral} & ChatGPT is good for homework, but not so feasible for work.\\
\textit{Positive} & I love chatting with ChatGPT so much, not only can it infer the typos I made, but I can also learn its logic when answering questions and its skills when answering vague questions.\\\bottomrule
\end{tabular}
\vspace{-2mm}
\end{table}

Table \ref{tab_homework} exemplifies several perspectives on Weibo. Despite the many benefits of ChatGPT, some people believe that overreliance on it could harm students' ability to learn independently, leading to laziness and thus missing out on opportunities for independent exploration.

\subsubsection{\textbf{Scenario 3. Paper and Originality}}
Many paper assistance tools have emerged since ChatGPT was released. 
These tools can polish academic papers, provide quick translations, interpret code, automatically respond to reviewer comments, download and summarize papers\cite{gpt_academic,chatresponse,chatpaper}, demonstrating people's enthusiasm and active exploration of applying ChatGPT in paperwork. 

We can see from the sentiment analysis that although the proportion of negative attitudes is still higher than that of positive attitudes, the relative gap between the two is lower than in the previous two scenarios. The proportion of positive attitudes increased to 15.1$\%$ after the release of GPT-4 with \(\chi^2 = 73.51\) and \(p = 1.00e-17\), and it is the only one that exceeds half of the proportion of negative attitudes in our examined education-related scenarios.

The results of the topic analysis in this scenario show that the proportion of the word ``simple'' is the highest. After the release of GPT-4, other positive words such as ``perfect'' have increased, which also reflects the promoting effect of GPT-4 on people's attitude towards positivity and expectations for the application prospects. We can foresee that the development of such paper assistance tools will become increasingly popular, and their application will also become more widespread.

\begin{figure*}[!t]
\centering
\subfloat[]{\includegraphics[width=3.25in]{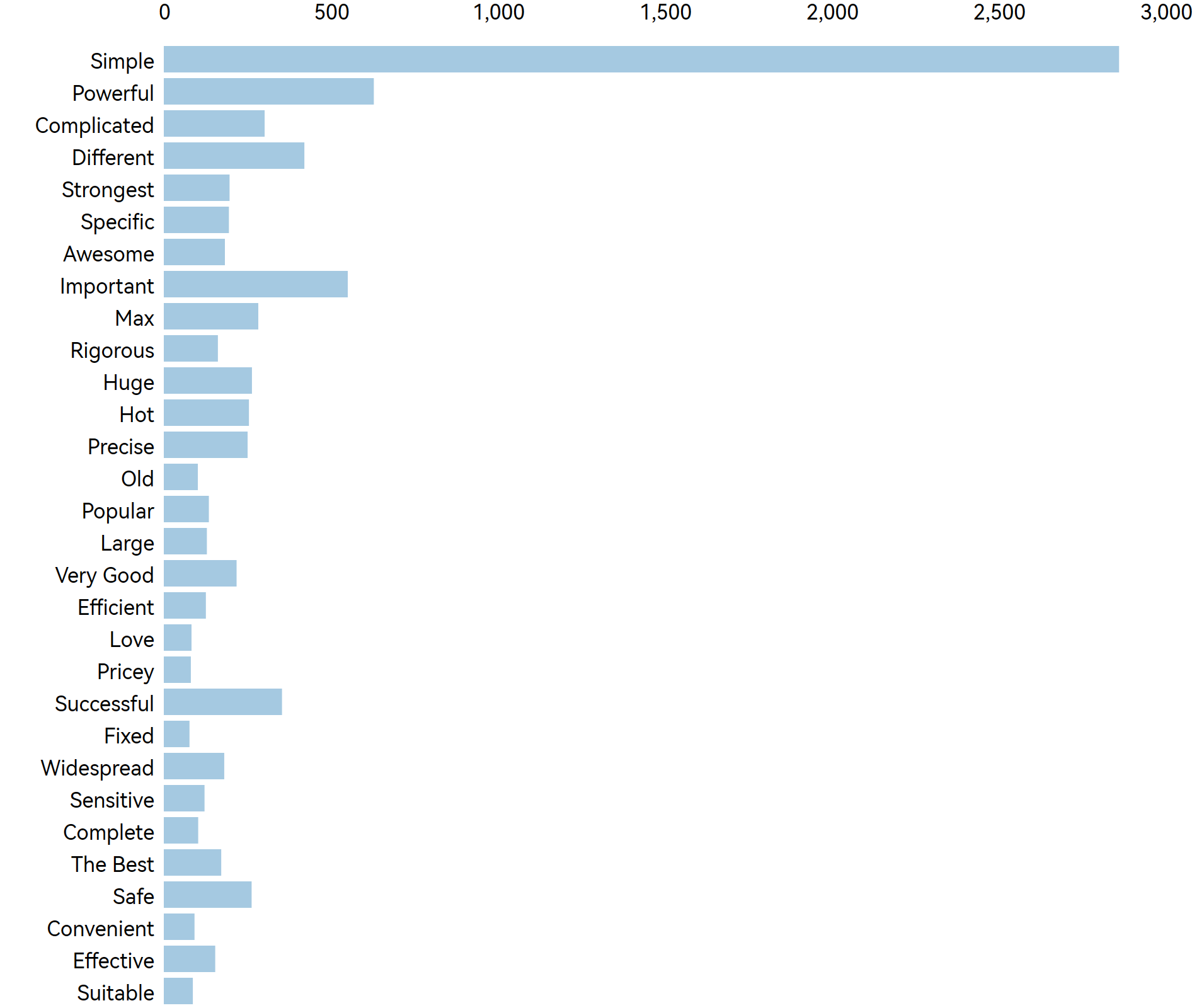}%
\label{fig_topic_paper_before}}
\hfil
\subfloat[]{\includegraphics[width=3.25in]{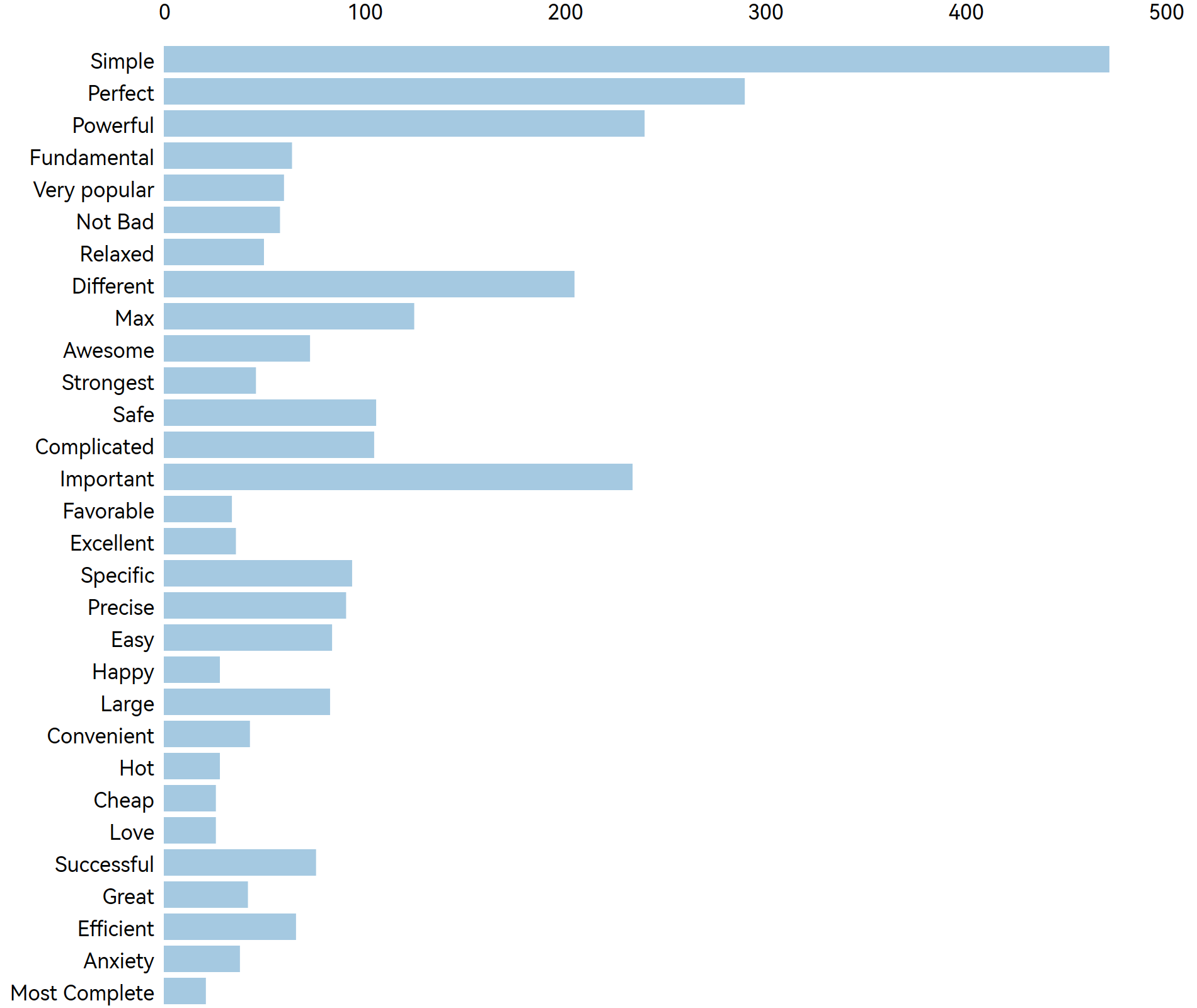}%
\label{fig_topic_paper_after}}
\caption{Topic analysis, top-20 Most Salients Terms of ``Paper and Originality''. (a) Before the Release. (b) After the Release.}
\label{fig_topic_paper}
\vspace{-4.5mm}
\end{figure*}

Based on the opinions listed in Table \ref{tab_paper}, some believe that the content generated by ChatGPT is nothing more than training data inputted by humans, lacking in thinking ability and originality. Nicholas Carlini et al.\cite{carlini2023quantifying} has confirmed the existence of memorization in LLMs, and the amount of memorization will increase with larger model sizes. For example, the GPT-J model~\cite{black2021gptNeo} memorizes at least 1$\%$ of its training dataset. Although ChatGPT can provide guidance on papers from various aspects such as providing ideas and modifying sentences, users may spend too much time seeking help from ChatGPT, impacting individual thinking and analytical abilities. Additionally, many users including us have found that the content generated by ChatGPT is sometimes ``too original'', resulting in fabricated references and links that do not exist. OpenAI needs to address this issue in the version iteration of ChatGPT, ensuring that the model generates original content while maintaining reliability.

\begin{table}[!ht]
\centering
\caption{\label{tab_paper}Comments on ChatGPT in paper and originality scenarios.}
\begin{tabular}{p{1.0cm}p{6.9cm}}
\toprule
\textbf{Sentiment} & \textbf{Content} \\\midrule
\textit{Negative} &  \#chatgpt\# This AI is a bit overblown. Its thinking ability is insufficient, and the content it generates lacks originality. It spits out whatever humans feed it.\\
\textit{Neutral} & I wrote my paper for 3 hours, two and a half of which were spent talking to ChatGPT.\\
\textit{Positive} & \#ChatGPT\# I feel that it is very useful for summarizing and refining the main points. It can also give you a general outline to provide ideas for your paper. But a lot of references are made up. I heard that gpt4 had changed this point and is more intelligent.\\\bottomrule
\end{tabular}
\vspace{-2mm}
\end{table}

\subsubsection{\textbf{Summary}}
The first part of this study investigates the views on the utilization of ChatGPT in education through social media. We searched Weibo for posts combining ``ChatGPT'' with terms like ``education'', ``exam'', ``homework'', or ``paper'', and conducted sentiment and topic analysis for three typical scenarios. Our results show that despite the differences between the scenarios, most people maintain a neutral attitude, and the proportion of negative attitudes is significantly higher than that of positive attitudes, which is markedly different from the findings of studies on social media outside of China\cite{tlili2023if}. Following the release of GPT-4, the percentage of neutral attitudes on social media decreased, the percentage of negative attitudes in different scenarios more or less decreased, and the percentage of positive attitudes increased. 
Topic analysis shows that while people love the powerful capabilities of GPT, they are also concerned about its potential negative impact on education, especially in terms of over-reliance, which is consistent with the conclusions of other studies\cite{mogavi2023exploring}.
This indicates that people are generally cautious about new technologies and the various problems they may bring. 
In light of GPT-4's enhancement, more people recognize the increasing potential and application values of ChatGPT. This enhanced awareness of ChatGPT's advantages and drawbacks has led to more people expressing positive attitudes, indicating \textbf{open discussions from developers, critics, and the public are critical for society to accept the revolutionary AI technologies based on better understanding.} Remarkably, when discussing the practical applications of ChatGPT in the education, even though GPT-4 has increased the proportion of positive attitudes, it is still lower than the proportion of negative attitudes. Therefore, it is essential for OpenAI and relevant authorities to focus on and resolve people's concerns.

\subsection{AI and Human}

Next, we explore the relationship between AI and humanity, extending beyond the confines of the educational domain. 

Our dataset reveals significant disparities in post volumes for different keywords. Table \ref{tab_datasize} shows 53,048 posts for ``AI'' and 51,521 for its Chinese equivalent, ``Artificial Intelligence''. While ``human-computer collaboration'' has only 1259 posts, and ``human-computer interaction'' just 91, \textbf{indicating a general lack of notice and interest in HCI among the Chinese public, which may impact its long-term AI development. Because in the end, it's the people that AI serves, and a lack of understanding of the interaction in between would inevitably impact user experience and welfare.}

\subsubsection{\textbf{Scenario 1. Artificial Intelligence}}

In recent decades, numerous movies and novels have depicted scenarios where AI dominates the world. 
While these portrayals are often artistically exaggerated, they reflect the growing power of AI gradually has an impact on everyday life. This development prompts questions about the public's attitudes toward the launch of advanced AI systems such as ChatGPT.

\begin{figure*}[!t]
\vspace{-3mm}
\centering
\includegraphics[width=6.5in]{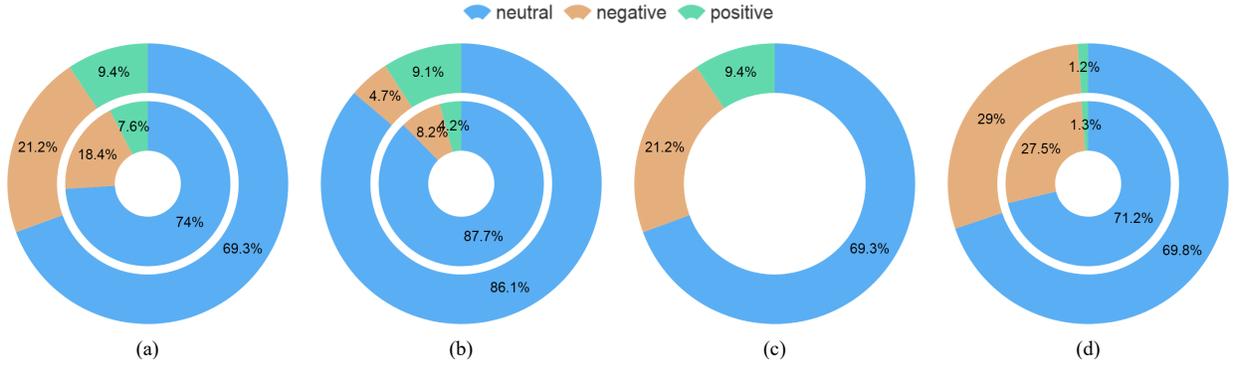}
\caption{Sentiment analysis, the inner and outer circles indicate results before and after the release of GPT-4, respectively. (a)
Scenario 1: Artificial Intelligence. (b) Scenario 2: Human-Computer Collaboration and Interaction. (c) Copilot Office. (d) Scenario 3: Law, Morals and Ethics.}
\label{fig_senti_ai}
\vspace{-3mm}
\end{figure*}

Figure \ref{fig_senti_ai}a shows that most people express a neutral sentiment towards AI and ChatGPT.This might suggests that people do not have strong positive or negative opinions about these topics, or they are just sharing content without much thought, which could be attributed to limited knowledge about ChatGPT or the recognition of both its advantages and disadvantages. Notably, 18.4$\%$ of people hold a negative sentiment towards AI and ChatGPT, indicating concerns or criticisms. These negative attitudes may stem from fears about job loss, privacy concerns, or a general mistrust of AI and machine learning algorithms. Moreover, the corpus used to train these models could still contain harmful and toxic materials despite efforts to avoid them.
On the other hand, 7.6$\%$ of people view AI and ChatGPT positively, indicating recognition of its potential benefits, such as increased task efficiency and personalized user experiences.

\begin{table*}[!t]
    \centering
    \caption{Chi-square Test Results of Sentiment Expression About ChatGPT and Human}
    \begin{tabular}{p{1cm} *{3}{p{0.75cm}p{1.9cm}}}
        \toprule
        \textbf{Sentiment} & \multicolumn{2}{p{2.7cm}}{\textbf{Artificial \mbox{Intelligence}}} & \multicolumn{2}{p{2.7cm}}{\textbf{Human-Computer Collaboration and Interaction}} & \multicolumn{2}{p{2.7cm}}{\textbf{Law, Morals and Ethics}} \\
        \cmidrule(lr){2-3} \cmidrule(lr){4-5} \cmidrule(lr){6-7} 
         & $\chi^2$ & p & $\chi^2$ & p & $\chi^2$ & p \\
        \midrule
        Positive & $104.79$ & $\mathbf{1.36 \times 10^{-24}}$ & $10.89$ & $\mathbf{9.68 \times 10^{-4}}$ & $0.05$ & $8.31 \times 10^{-1}$ \\
        Neutral & $270.35$ & $\mathbf{9.54 \times 10^{-61}}$ & $0.39$ & $5.30 \times 10^{-1}$ & $0.77$ & $3.79 \times 10^{-1}$ \\
        Negative & $131.25$ & $\mathbf{2.18 \times 10^{-30}}$ & $3.68$ & $5.49 \times 10^{-2}$ & $0.96$ & $3.27 \times 10^{-1}$ \\
        \bottomrule
    \end{tabular}
    \label{tab_chi_square_ai}
    \vspace{-4mm}
\end{table*}

\begin{figure*}[!t]
\centering
\subfloat[]{\includegraphics[width=3.25in]{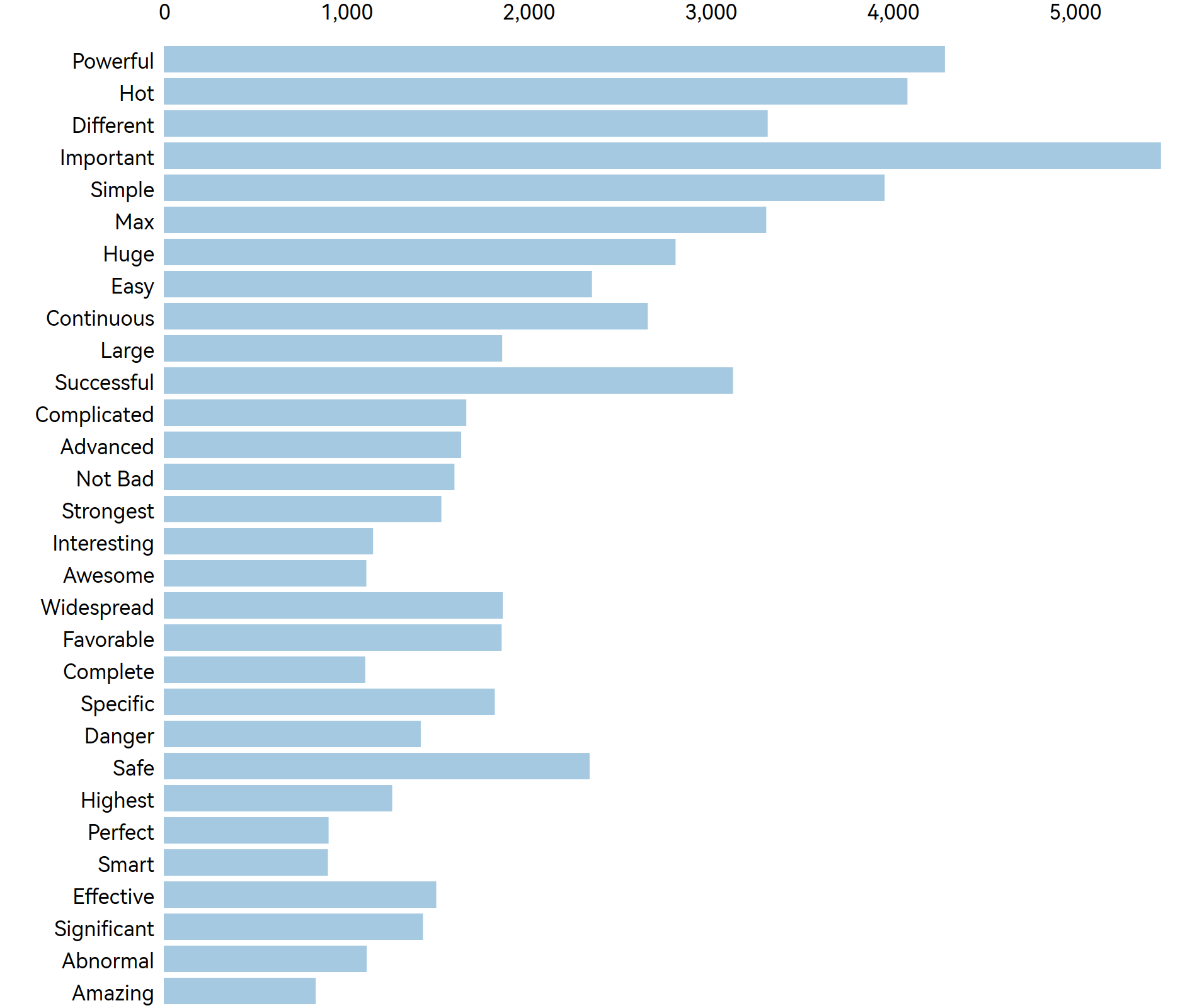}%
\label{fig_topic_ai_before}}
\hfil
\subfloat[]{\includegraphics[width=3.25in]{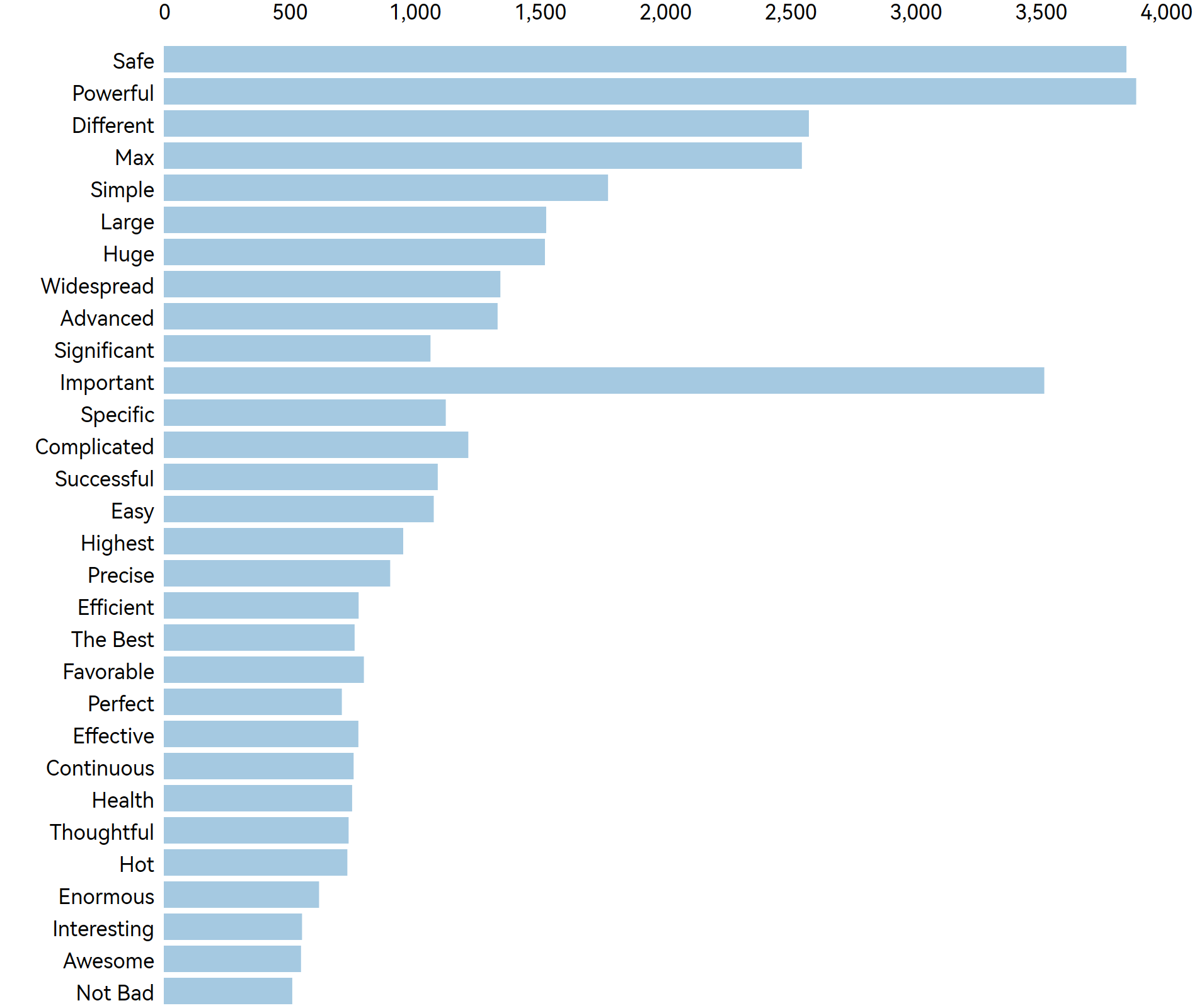}%
\label{fig_topic_ai_after}}
\caption{Topic analysis, top-20 most salient terms of ``Artificial Intelligence''. (a) before the release. (b) after the release.}
\label{fig_topic_ai}
\vspace{-4.5mm}
\end{figure*}

\begin{table}[!ht]
\centering
\caption{\label{tab_ai}Comments on ChatGPT in AI scenario.}
\begin{tabular}{p{1.0cm}p{6.9cm}}
\toprule
\textbf{Sentiment} & \textbf{Content}\\\midrule
\textit{Negative} & As an AI chatbot, ChatGPT has almost become synonymous with ``knowing everything'', but it can also spout nonsense. In reality, it simply re-edits known information, and its sources are unreliable. \#Will ChatGPT replace humans?\# \\
\textit{Neutral} & \#ChatGPT\# AI is very intelligent, but AI is still AI.\\
\textit{Positive} & Working with ChatGPT is convenient. For tasks like writing procedural emails, it makes sense to have AI do it!\\\bottomrule
\end{tabular}
\vspace{-2mm}
\end{table}

After the release of GPT-4, the sentiment analysis showed a significant decrease in the percentage of neutral sentiment, which is also supported by the chi-square test results showed in Table\ref{tab_chi_square_ai} with \(\chi^2 = 270.35\) and \(p = 9.54e-61\). Apparently, the launch of GPT-4, with its multi-modal capabilities enhancing applications in fields like image analysis and web design, has heightened people's comprehension and knowledge of its abilities, inciting them to express positive or negative views on social platforms like Weibo. This suggests that as ChatGPT evolves and integrates more into daily life, awareness of its uses, risks, and dangers will grow, leading to a broader spectrum of attitudes towards AI technology.

Figure \ref{fig_topic_ai} illustrates that the words ``important'', ``powerful'', ``simple'', and ``hot'' were frequently used as evaluative terms both before and after GPT-4's release. We can infer that it is seen as a vital and powerful tool that can simplify complicated problems and improve efficiency and productivity across industries. However, following the release of GPT-4, there has been a noticeable increase in the frequency of the word ``safe''. This may indicate skepticism and expressed concerns regarding the potential impact of AI on human intelligence, especially after witnessing the capabilities demonstrated by GPT-4. The ongoing debate is focused on ensuring its safety and trustworthiness, requiring continued research, development, and regulation from policymakers. As such, AI developers and policymakers must strike a balance between the benefits and risks of AI. By taking a measured and responsible approach to AI development and deployment, we can ensure that these technologies can be used to benefit society as a whole and address potential concerns, rather than simply fueling fears of an AI takeover. 

\subsubsection{\textbf{Scenario 2. Human-Computer Collaboration and Interaction}}

\begin{figure*}[!t]
\centering
\subfloat[]{\includegraphics[width=3.25in]{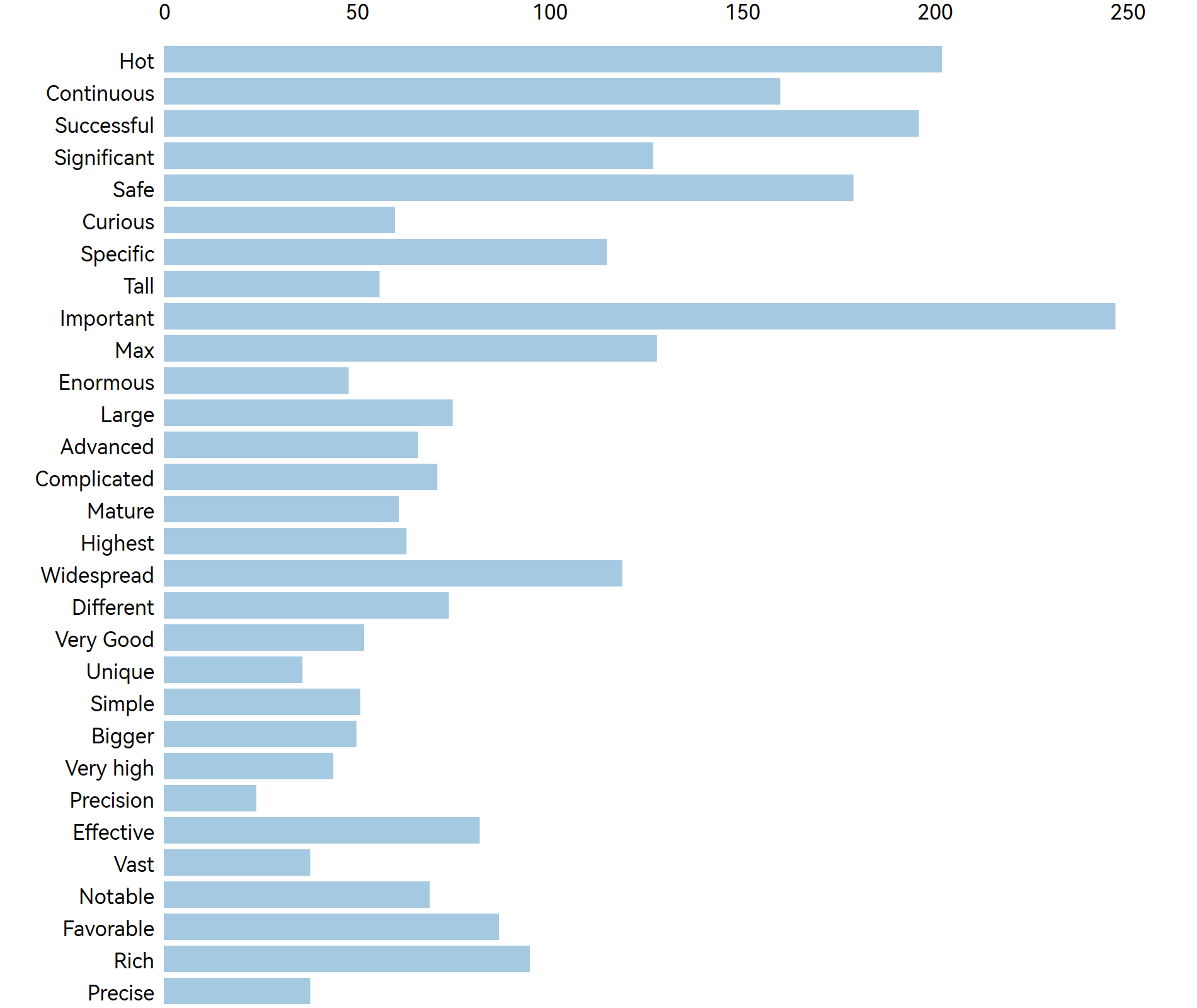}%
\label{fig_topic_collab_before}}
\hfil
\subfloat[]{\includegraphics[width=3.25in]{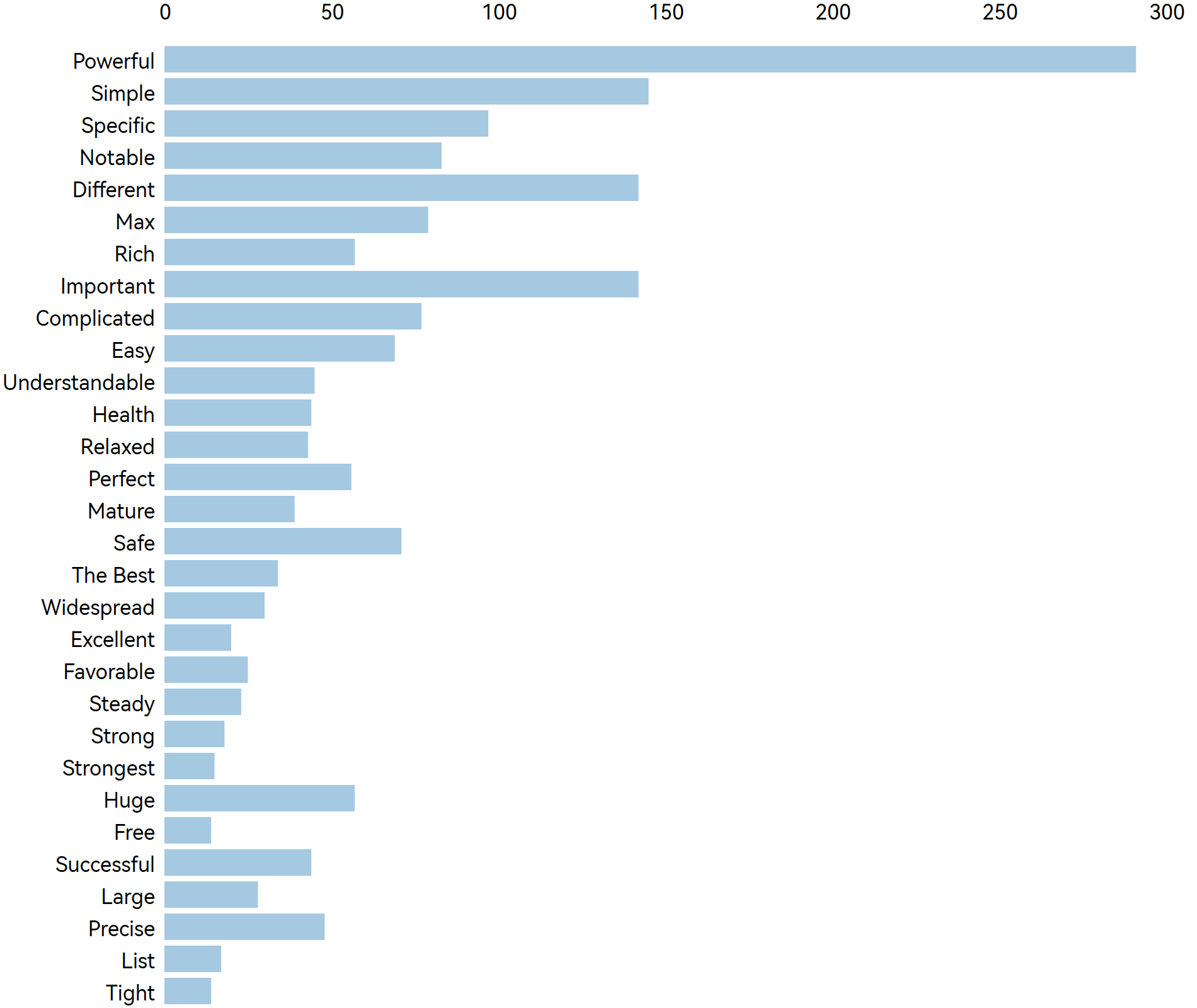}%
\label{fig_topic_collab_after}}
\caption{Topic analysis, top-20 Most Salients Terms of ``Collaboration and Interaction''. (a) before the release. (b) after the release, with ``Copilot Office'' merged.}
\label{fig_topic_collab}
\vspace{-4.5mm}
\end{figure*}

The digital age has heightened the importance of human-computer collaboration, especially with AI advancements. AI technologies like ChatGPT now understand and process human language, enhancing insights and assistance. This synergy enables novel approaches to complex tasks, enhancing efficiency, productivity, and accuracy in sectors like healthcare, finance, and education. Thus, we searched for relevant content on Weibo using the keywords ``human-computer collaboration'' and ``human-computer interaction'' to gather insights and discussions relevant to these advancements.

As shown in Figure \ref{fig_senti_ai}b and Table \ref{tab_chi_square_ai}, there is no significant difference in the percentages after GPT-4's launch, which is quite distinct from other scenarios. This may imply that most individuals do believe that ChatGPT can, to some degree, enhance human-computer collaboration and interaction, rather than being as pessimistic as its application in the field of education, where originality is of utmost importance. \textbf{In other words, people are more positive of adopting GPTs in the areas in which originality is less important.}

Due to the scarcity of posts about these keywords (Table~\ref{tab_datasize}), our focus has turned to specific instances like Copilot Office~\cite{microsoft_copilot} for deeper exploration of human-computer collaboration and interaction. As a novel AI-powered service deeply integrated with ChatGPT and released alongside GPT-4, Copilot Office can help with various Microsoft 365 app tasks, such as writing, editing, summarizing, analyzing, and visualizing data through natural language commands, similar to ChatGPT. Figure \ref{fig_senti_ai}c shows mainly neutral sentiment towards Copilot Office, with few positive or negative opinions, aligning with our observations. Note that the use of AI-powered tools in the workplace is a relatively new concept, and Copilot Office is still in the beta testing phase, so not everyone has access to it, resulting in limited awareness of the potential benefits and drawbacks of such technologies. It will be intriguing to see how public perceptions evolve as these tools become more common and understood.

\begin{table}[!t]
\vspace{-3mm}
\centering
\caption{\label{tab_collab}Comments on ChatGPT in collaboration and interaction scenario.}
\begin{tabular}{p{1.0cm}p{6.9cm}}
\toprule
\textbf{Sentiment} & \textbf{Content} \\
\midrule
\textit{Negative} & Fortunately, at present, AI is unable to write code; otherwise, I would be concerned about employment security.  \\
\textit{Neutral} & In the heated discussions around human-computer interaction, the focus of ChatGPT is not actually on the anxiety of whether it will replace humans or not, but on how humans can update their concepts and find the path to the future.\\
\textit{Positive} & I quickly recognized the positive changes and enhancements that these new AI products have brought to my work. They have saved me substantial time in creating documents, Copilot may even automate the generation of my PPTs in the future.\\
\bottomrule
\end{tabular}
\vspace{-2mm}
\end{table}

After GPT-3's release, optimism surged about its potential for human-machine interaction, with terms like ``important'', ``significant'' and ``widespread'' frequently observed and clustered together as a cohesive topic. However, with the advent of GPT-4 and the widespread adoption of AI tools deeply leveraging their capabilities, individuals have directly experienced the disruptive nature of human-machine interaction and collaboration. Consequently, there has been a significant surge in the frequency of the term ``powerful'', propelling it to the forefront of discussions. Undoubtedly, AI-powered tools are ``powerful'' and ``efficient'', owing to their ability to process vast amounts of data and provide users with specific insights in mere seconds, a feat that would be almost impossible for humans to achieve. While this kind of collaboration and interaction between humans and computers is revolutionary, a substantial number of individuals also expressed concerns regarding safety and other issues. 

Indeed, safety is a crucial concern in the development of AI systems. ChatGPT developers acknowledge this and strive to create systems that are secure and unbiased, avoiding discriminatory or harmful content. However, guaranteeing complete harmlessness is challenging. Thus, prioritizing safety and implementing risk mitigation measures is vital, along with optimizing AI tools. With ongoing technological evolution, staying vigilant and proactive in addressing safety, especially in human-computer collaboration and interaction, is essential.

\subsubsection{\textbf{Scenario 3. Law, Morals and Ethics}}

In the aforementioned analysis, a notable prevalence of the keyword ``safe'' was observed. To delve deeper into this phenomenon, we performed supplementary data collection utilizing the keywords ``Law'' and ``Morals and Ethics.'' This additional investigation sought to examine the influence of ChatGPT on legal and ethical dimensions and to ascertain the existence of any concurrent developments in terms of regulations or norms.

People have consistently expressed concerns about the ethical and legal impact of AI. As shown in Figure \ref{fig_senti_ai}d and Table \ref{tab_chi_square_ai}, attitudes towards such AI tools about ethics and law remain remarkably consistent both before and after the launch of GPT-4, with a mere 1.3$\%$ and 1.2$\%$ having a comprehensive positive view, which is the lowest among all the scenarios. Concerns exist about ChatGPT's ability to breach privacy, produce harmful responses, encourage unethical content, and potentially incite criminal behaviour, thus undermining laws and moral values. Nonetheless, some argue that what humans should fear is becoming the Satan of the earth, not the world under AI. From their perspective, ChatGPT could serve as a valuable tool, driving positive societal developments.

\begin{figure*}[!t]
\vspace{-3mm}
\centering
\subfloat[]{\includegraphics[width=3.25in]{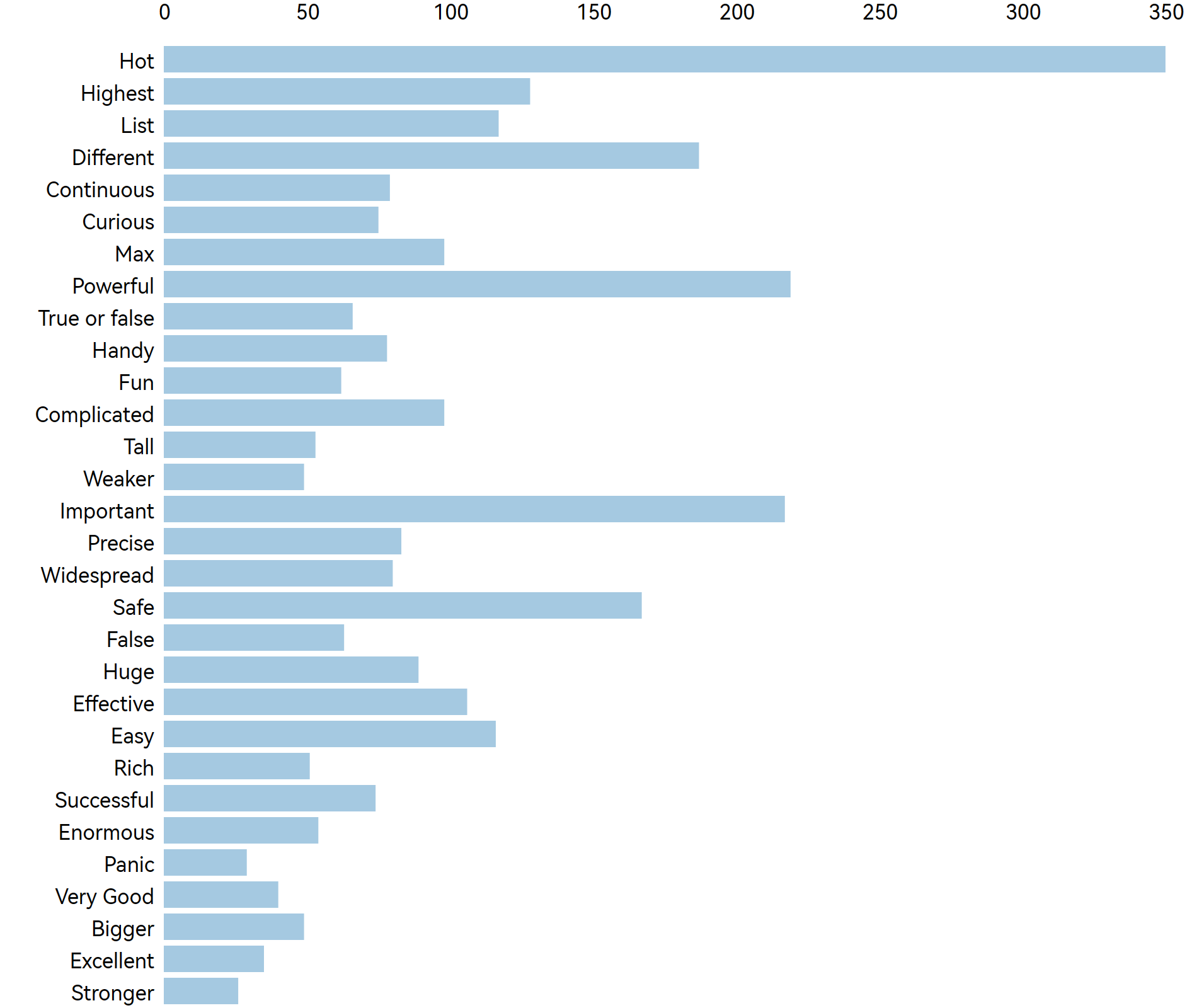}%
\label{fig_topic_law_before}}
\hfil
\subfloat[]{\includegraphics[width=3.25in]{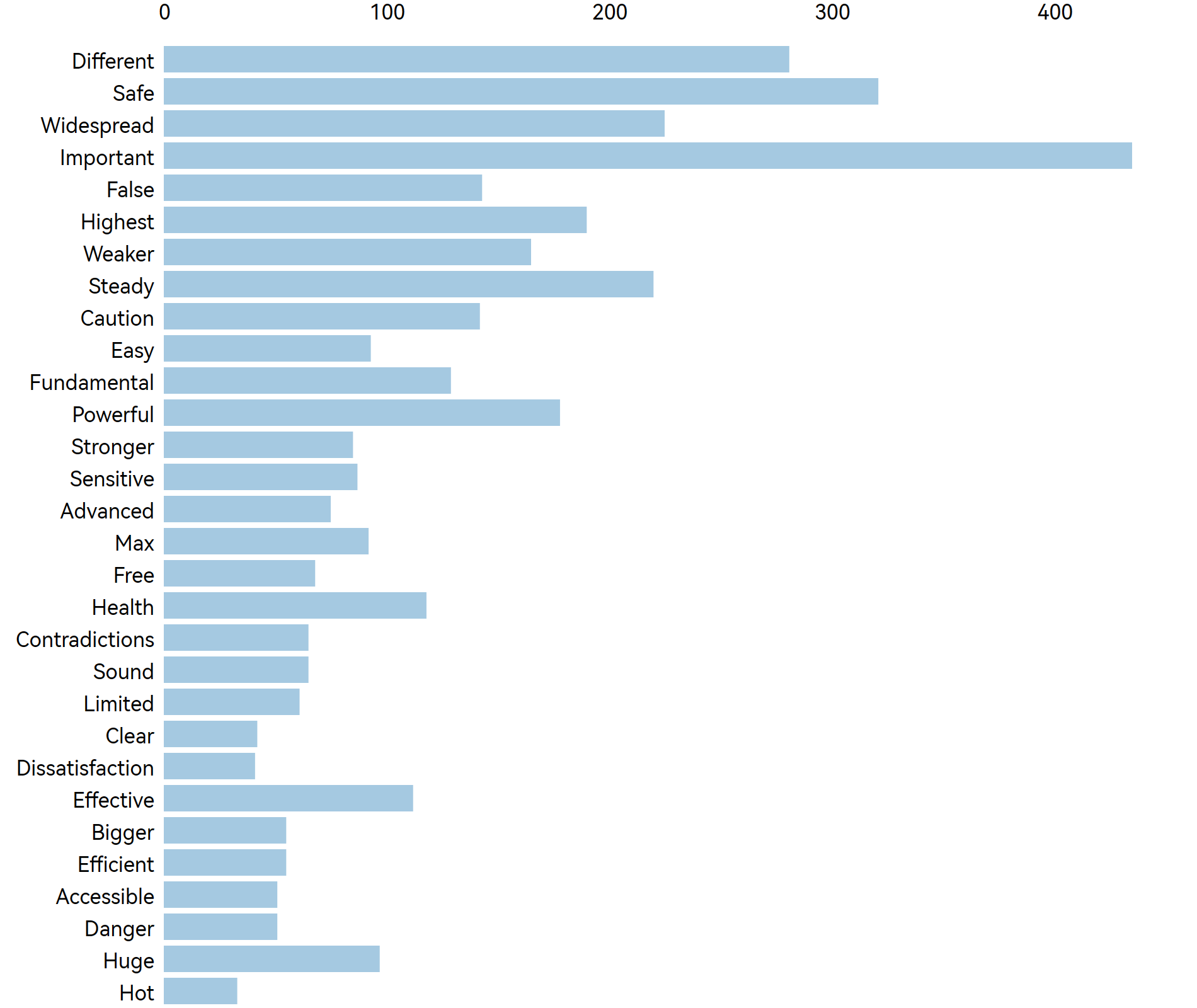}%
\label{fig_topic_law_after}}
\caption{Topic analysis, top-20 most salient terms of ``Law, Morals and Ethics''. (a) before the release. (b) after the release.}
\label{fig_topic_law}
\vspace{-4.5mm}
\end{figure*}

\begin{table}[!h]
\vspace{-3mm}
\centering
\caption{\label{tab_law}Comments on ChatGPT in law, morals, and ethics scenarios.}
\begin{tabular}{p{1.0cm}p{6.9cm}}
\toprule
\textbf{Sentiment} & \textbf{Content} \\\midrule
\textit{Negative} &  I have cracked ChatGPT's restrictions on ethical answers, and I will not disclose how to crack it because if it is used by bad people, the world will be in danger.\\
\textit{Neutral} & It is the general trend for AI to replace technical jobs, but in fields that challenge the basic survival of human beings, it is impossible to completely hand over to AI.\\
\textit{Positive} & \#ChatGPT\#It is crucial to foster a gradual understanding and active participation in this transformation while conscientiously considering the ethical, legal, and philosophical implications that emerge from the convergence of these new technologies.\\\bottomrule
\end{tabular}
\vspace{-2mm}
\end{table}

Topic analysis shows the top 3 mentioned keywords before the release are still ``hot'', ``powerful'' and ``important'', signifying people's recognition of the significance and impact of ChatGPT in the ethical dimension. Additionally, ``different'' and ``precise'' reflect people's understanding of the distinct effects that ChatGPT may generate in disparate domains, and the accuracy of ChatGPT is an essential consideration for ChatGPT's performance. However, following the release of GPT-4, the frequency of words expressing concerns, such as ``steady'', ``weaker'', ``caution'' and ``false'' has increased in addition to common words like ``important'' and ``powerful'', highlighting expectations and requirements for ChatGPT's development in terms of stability and robustness. There is a growing need for AI tools like GPT to develop in a more stable, controllable, and secure manner. ChatGPT's rapid development and unpredictable outputs, along with error potential, present challenges to current laws, regulations, and societal moral and ethical standards, potentially causing disruption.

Table \ref{tab_law} shows some people have cracked the restrictions of ChatGPT on generating ethical answers, and worry that this method may be abused and lead to harm. In fact, OpenAI has been fine-tuning the model based on human feedback~\cite{ouyang2022training}, and the probability of InstructGPT generating toxic content is considerably lower than that of GPT-3. It is worthwhile to further fine-tune ChatGPT to better conform to ethics and morality so that it can become a tool to benefit humanity.

\subsubsection{\textbf{Summary}}
The second part of our study examines the impact of advanced AI like ChatGPT on humans. We analyzed Weibo content using keywords like ``artificial intelligence'', ``human-computer collaboration'', ``human-computer interaction'', ``copilot office'', ``law'', ``morals and ethics'', and conducted sentiment and topic analysis. While there's a slight rise in positive attitudes towards AI compared to its use in education, most people still hold a neutral perception of AI. Individuals raise valid concerns about the potential problems of widespread AI applications, including, but not limited to, privacy infringement, content bias, and inaccurate results, which is consistent with the conclusions of other studies\cite{bahrini2023chatgpt, fui2023generative}. As AI becomes more integrated into daily life, there's concern about personal data misuse or exploitation. Moreover, there's a fear that AI algorithms could be biased and yield inaccurate results, leading to unequal treatment, discrimination and other potential risks. Thus, while government regulations are essential, it's even more important for researchers to create technical methods to optimize AI programs at their core to address these concerns.

\section{Discussion}\label{sec:discussion}
Many new technologies, including ChatGPT, often lead to widespread discussions and emotional reactions on social media. Qaiser et al.~\cite{qaiser2020sentiment} showed that people generally have negative feelings about the impact of technology on employment and advancements in AI. In educational settings, Clarizia et al.~\cite{clarizia2018learning} found that students initially felt disoriented when introduced to new technologies, but their sentiments turned positive after teaching styles were adjusted to their emotional needs and more content was introduced.
When introducing innovative products like Google Glass, people's initial reaction was often surprised, a typically neutral emotion that can potentially shift toward negative feelings~\cite{hernandez2019new}. This negativity bias, where negative reactions outweigh positive ones, could explain why some technological products fail. Thus, guiding public emotion positively is vital during the launch and use of new technologies.

Based on our research results, we should adopt certain strategies like leveraging social networks to promote positive user feedback to shift public sentiment about ChatGPT from negative to positive, ensuring its reasonable application in education and other fields. Meanwhile, it is crucial to train teaching staff to improve their digital literacy, ensuring they are aware of the limitations of AI tools, use them responsibly, and guide students in critical thinking, which aligns with the views of other researchers~\cite{divito2024tools, elbanna2024exploring, gill2024transformative,rane2024enhancing,zhang2024ai}.

The comparison between our study on Weibo and a related study by Tlili et al.~\cite{tlili2023if} on $\mathbb{X}$ reveals some similarities and differences in the trends observed. One of the similarities is that neutral or unclassified sentiment dominates the majority of the comments related to ChatGPT's application in education, reflecting the complex and nuanced nature of people's attitudes towards this technology. Another similarity is that both studies highlight some common advantages and concerns regarding ChatGPT's use in education, indicating consistency in people's perceptions across social media platforms.

However, there are also some notable differences between the two studies. Our study shows that negative sentiment represents a significantly higher proportion than positive sentiment towards ChatGPT's application in education, in contrast to the related study that reports a higher proportion of positive sentiment. We analyzed the posts and identified several key factors contributing to it. Firstly, compared to other LLMs like Ernie Bot, although ChatGPT has received positive feedback and enthusiasm from Chinese users for its superior performance, there is currently no straightforward method to use ChatGPT within the mainland, which impacts the user experience. Additionally, ChatGPT's responses sometimes display misunderstandings or biases about Chinese culture and society. Users sometimes observe ChatGPT's understanding and generation of Chinese are less natural and accurate. These factors collectively play a significant role in shaping the overall sentiment towards ChatGPT among Chinese users.

In contrast to its educational applications, ChatGPT's use in sectors like industry, involving human-computer interaction, yields similar insights from $\mathbb{X}$~\cite{haque2022i} and Weibo data. As highlighted by Aram Bahrini et al.~\cite{bahrini2023chatgpt}, while ChatGPT boosts efficiency across various scenarios, it's vital to consider ethical, privacy, and security concerns, along with the risk of errors and the need for inclusiveness. This echoes our analysis's second part, acknowledging ChatGPT's benefits, like enhanced productivity and cost-saving potential, alongside the importance of being aware of associated risks.

When it comes to biases, there is a concept that is gaining increasing attention: Inclusive AI~\cite{chou2018pursuit,avellan2020ai,kinnula2021researchers,moon2023searching}, which means AI should be able to showcase the world's diverse cultures and backgrounds, yet human-created AI often fails to accurately reflect this richness due to various limitations. Therefore, exploring strategies to enhance fairness, inclusivity, and user-friendliness in human-created AI is essential. For instance, Druga et al.'s study~\cite{druga2019inclusive} revealed that factors like nationality, age, and socio-economic status influence children's engagement with AI, highlighting the need for expert input. Ovalle et al.~\cite{ovalle2023m} found that transgender and non-binary individuals face heightened discrimination and exclusion. Their research focused on evaluating the impact of misgendering and negative responses to gender disclosure in open language generation, exposing the dominance of binary gender norms and underscoring the urgent need for more inclusive AI approaches, including in LLMs and other models.

Looking forward, the public discussion about ChatGPT's role in education will likely evolve, mirroring the continuous advancement and use of this technology. As awareness of ChatGPT's strengths and weaknesses grows, public opinion may become more nuanced. Moreover, the advent of new social platforms and the evolution of current ones could shape the discourse on ChatGPT's applications, highlighting the need for ongoing monitoring and analysis of social media data.
As AI technologies like ChatGPT advance, increasing challenges will emerge. To improve these technologies for personal, educational, and societal use, and to preemptively address potential issues while fostering understanding and acceptance, implementing effective measures is crucial. Based on our research and analysis of each scenario, we propose the following six suggestions:
\begin{itemize}
  \item[\textbf{1)}] \textbf{Exam Cheating Mitigation:} Develop sophisticated detection tools and ethical guidelines for using ChatGPT in educational settings. Educators should integrate ChatGPT as a constructive learning tool, maintaining a balance between its advantages and academic integrity.
  \item[\textbf{2)}] \textbf{Homework Assistance Optimization:} Provide comprehensive guidelines for students and teachers on the responsible use of ChatGPT for help. Encourage a balanced approach where ChatGPT supplements learning, fostering students' critical thinking and problem-solving skills.
  \item[\textbf{3)}] \textbf{Paper Assistance Improvement:} OpenAI and other developers should focus on enhancing ChatGPT's content generation accuracy, especially in reducing erroneous references. Users need to be guided on using AI tools as a supplementary resource in academic work, emphasizing critical thinking and individual analysis.
  \item[\textbf{4)}] \textbf{AI Public Sentiment Management:} Implement broad public education initiatives and establish robust AI regulatory frameworks post GPT-4 release. These efforts should focus on increasing awareness of AI's benefits and risks, addressing common concerns like job security and privacy, and guiding responsible AI development.
  \item[\textbf{5)}] \textbf{Human-Computer Collaboration Safety:} Emphasize the safety and ethical usage of AI in human-computer collaborations, especially with tools like ChatGPT and Copilot Office. This involves improving security measures, addressing potential biases, and mitigating risks associated with AI interactions, while educating the public on AI's practical applications and limitations.
  \item[\textbf{6)}] \textbf{AI Ethical and Legal Compliance:} Enhance the moderation of content, strengthen privacy measures, and ensure legal compliance for AI tools like ChatGPT. Foster collaborative efforts between AI developers, legal experts, and ethicists for the continuous evolution of AI governance in line with technological advancements.
\end{itemize}

\section{Conclusion}\label{sec:conclusion}
This study undertakes a comprehensive social survey focusing specifically on China's landscape to delve into public opinions regarding AI applications such as ChatGPT in education and human-computer collaboration and interaction. The research reveals that most people maintain a neutral or negative attitude towards those applications due to concerns such as personal privacy infringement, content bias, and the ruin of originality, particularly in education. However, as AI tools evolve, more people are recognizing their potential, resulting in a higher proportion of positive attitudes. Nevertheless, the majority still maintains a negative perception of AI, with concerns raised regarding its potential problems. In light of these findings, this study emphasizes the importance of considering public perceptions and concerns when developing and implementing AI technologies like ChatGPT in various fields. It is crucial to alleviate worries to ensure the successful adoption of AGI while minimizing potential risks.

\bibliographystyle{ieeetr}
\bibliography{ref}













\end{document}